\documentclass[apj,twocolappendix,iop]{emulateapj}
\usepackage{graphicx}
\usepackage{color}
\usepackage{xspace}
\usepackage{amsmath}
\usepackage{amssymb}
\usepackage{ifthen}
\usepackage[nodayofweek]{datetime}
\usepackage{hyperref} 

\if {a}{b} \usepackage[authoryear]{natbib} \fi

\newcommand{\Mpc}{\,{\rm Mpc}}

\newcommand{\Msun}{\,{\rm M}_{\odot}}

\newcommand{\Msunyr}{\ensuremath{\,\Msun\, {\rm yr}^{-1}}}
\newcommand{\ang}{\ensuremath{\,\textrm{\AA}}\xspace}
\newcommand{\Mstellar}{\ensuremath{M_{*}}}

\newcommand{\ergsec}{\ensuremath{\,{\rm erg}\,\, {\rm s}^{-1}}}
\newcommand{\kms}{\ensuremath{\, {\rm km\, s^{-1}}}}

\def \apjs {ApJS}
\def \apjl {ApJL}
\def \aap {A\&A}

\newcommand{\Ha}{{\rm H}\ensuremath{\alpha}\xspace}
\newcommand{\Hb}{{\rm H}\ensuremath{\beta}\xspace}
\newcommand{\nii}{{\rm [N~\textsc{ii}]}}
\newcommand{\sii}{\rm [S~\textsc{ii}]}
\newcommand{\oi}{[\rm O~\textsc{i}]}
\newcommand{\oii}{\rm [O~\textsc{ii}]}
\newcommand{\oiii}{\rm [O~\textsc{iii}]}

\usepackage{boolexpr}
\newcommand{\n}[1]{%
  \switch[\pdfstrcmp{#1}]%
  \case{{nsample}}772% The sample size
  \case{{nsample0}}832% Initial sample size
  \case{{fibres}}819% Total fibres in SAMI
  \case{{specfwhmred}}${1.607}^{+0.075}_{-0.052}\ang$% From Jesse's SDN
  \case{{specfwhmblue}}${2.650}^{+0.122}_{-0.088}\ang$%
  \case{{specR-blue}}1812%
  \case{{specR-red}}4263%
  \case{{specDeltaSigma-blue}}\ensuremath{70\kms}%
  \case{{specDeltaSigma-red}}\ensuremath{30\kms}%
  \case{{meanSeeing}}\ensuremath{2.16}%
  \case{{meanSeeingStd}}0.41%
  \case{{lzifuInstSigRed}}${0.72\ang}$% From I-Ting via email 19 May 2016
  \case{{lzifuInstSigBlue}}${1.15\ang}$%
  \otherwise\textbf{\color{red}???}%
  \endswitch%
}

\usepackage{acronym}

\newcommand{\aafilvalue}[1]{%
  \switch[\pdfstrcmp{#1}]%
  \case{{AAO}}Australian Astronomical Observatory, 105 Delhi Rd, North Ryde, NSW 2113, Australia%
  \case{{SIfA}}Sydney Institute for Astronomy, School of Physics, A28, The University of Sydney, NSW, 2006, Australia%
  \case{{RSAA}}Research School of Astronomy and Astrophysics, Australian National University, Canberra, ACT 2611, Australia%
  \case{{IfA}}Institute for Astronomy, University of Hawaii, 2680 Woodlawn Drive, Honolulu, HI 96822, USA%
  \case{{CAASTRO}}ARC Centre of Excellence for All-Sky Astrophysics (CAASTRO)%
  \case{{UMelb}}School of Physics, The University of Melbourne, Parkville, VIC 3010, Australia%
  \case{{UQ}}School of Mathematics and Physics, University of Queensland, QLD 4072, Australia%
  \case{{MQ}}Department of Physics and Astronomy, Macquarie University, NSW 2109, Australia%
  \case{{Swin}}Swinburne University, Hawthorn, VIC 3122, Australia%
  \case{{UNSW}}School of Physics, University of New South Wales, NSW 2052, Australia%
  \case{{ICRAR}}International Centre for Radio Astronomy Research, University of Western Australia, 35 Stirling Highway, Crawley WA 6009, Australia%
  \otherwise\textbf{\color{red}???}%
  \endswitch%
}

\newcommand{\aafil}[1]{\affiliation{\aafilvalue{#1}}}

\makeatletter%
\@ifclassloaded{emulateapj}%
  {\journalinfo{Draft date \longdate{\today}}%
   \submitted{submitted to MNRAS on 17 May 2017}
   \shorttitle{SAMI Galaxy Survey: Data Release One}
   \shortauthors{Green et al.}%
  }
  {\relax}%
\makeatother%

\begin{document}
\title[The SAMI Galaxy Survey: Data Release One]{The SAMI Galaxy Survey: Data Release One with Emission-line Physics Value-Added Products}

% Acronyms appearing in the paper (try to keep this list short!)
\newacro{IFS}{integral-field spectroscopy}
\newacro{IFU}{integral-field unit}

\newcommand{\am}[1]{\relax}

% Authors up to date through Rob Sharp - AWG 27 Jan

\author{Andrew W. Green${}^{\dag}$\am{AAO}}
%\affiliation{Australian Astronomical Observatory, PO Box 915, North Ryde, NSW 1670, Australia}
\aafil{AAO}
\email{andrew.green@aao.gov.au}
\thanks{${}^{\dagger}$How each author contributed to the paper is listed at the end.}

\author{Scott M. Croom\am{Syd}\am{CO}}
\aafil{SIfA}
\aafil{CAASTRO}

\author{Nicholas Scott\am{Syd}}
\aafil{SIfA}

\author{Luca Cortese\am{ICRAR}}
\aafil{ICRAR}

\author{Anne M. Medling\am{RSAA}\am{CIT}\am{HF}}
\aafil{RSAA}
\affiliation{Cahill Center for Astronomy and Astrophysics, California Institute of Technology, MS 249-17, Pasadena, CA 91125, USA}
\affiliation{Hubble Fellow}

\author{Francesco D'Eugenio}
\aafil{CAASTRO}
\aafil{RSAA}

\author{Julia J. Bryant\am{AAO}\am{Syd}\am{CO}}
\aafil{AAO}
\aafil{SIfA}
\aafil{CAASTRO}

\author{Joss Bland-Hawthorn\am{Syd}}
\aafil{SIfA}

\author{J. T. Allen\am{Syd}\am{CO}}
\aafil{SIfA}
\aafil{CAASTRO}

\author{Rob Sharp}
\aafil{RSAA}

\author{I-Ting Ho\am{RSAA}\am{MPA}\am{IfA}}
%\aafil{RSAA}
\affiliation{Max Planck Institute for Astronomy, K\"{o}nigstuhl 17, 69117 Heidelberg, Germany}
%\aafil{IfA}

\author{Brent Groves\am{RSAA}}
\aafil{RSAA}

\author{Michael J. Drinkwater\am{UQ}\am{CO}}
\aafil{UQ}
\aafil{CAASTRO}

\author{Elizabeth Mannering}
\aafil{AAO}

\author{Lloyd Harischandra}
\aafil{AAO}

\author{Jesse van de Sande\am{Syd}}
\aafil{SIfA}

\author{Adam D. Thomas\am{RSAA}}
\aafil{RSAA}

\author{Simon O'Toole}
\aafil{AAO}

\author{Richard M. McDermid\am{AAO}}
\aafil{AAO}
\aafil{MQ}

\author{Minh Vuong}
\aafil{AAO}

\author{Katrina Sealey}
\aafil{AAO}

% Builders

\author{Amanda E. Bauer}
\aafil{AAO}

\author{S. Brough\am{AAO}}
\aafil{UNSW}

\author{Barbara Catinella\am{ICRAR}}
\aafil{ICRAR}

\author{Gerald Cecil\am{UNC}}
\affiliation{Dept. Physics and Astronomy, University of North Carolina, Chapel Hill, NC 27599, USA}

\author{Matthew Colless\am{RSAA}}
\aafil{RSAA}

\author{Warrick J. Couch}
\aafil{AAO}

\author{Simon P. Driver}
\aafil{ICRAR}

\author{Christoph Federrath}
\aafil{RSAA}

\author{Caroline Foster\am{AAO}}
\aafil{AAO}

\author{Michael Goodwin\am{AAO}}
\aafil{AAO}

\author{Elise J. Hampton\am{RSAA}}
\aafil{RSAA}

\author{A. M. Hopkins\am{AAO}}
\aafil{AAO}

\author{D. Heath Jones}
\affiliation{English Language and Foundation Studies Centre, University of Newcastle, Callaghan NSW 2308, Australia}

\author{Iraklis S. Konstantopoulos\am{AAO}\am{Atl}}
\aafil{AAO}
\affiliation{Atlassian 341 George St Sydney, NSW 2000, Australia}

\author{J.S. Lawrence\am{AAO}}
\aafil{AAO}

\author{Sergio G. Leon-Saval}
\aafil{SIfA}

\author{Jochen Liske\am{Ham}}
\affiliation{Hamburger Sternwarte, Universit{\"a}t Hamburg, Gojenbergsweg 112, 21029 Hamburg, Germany}

\author{\'Angel R. L\'opez-S\'anchez\am{AAO}}
\aafil{AAO}
\aafil{MQ}

\author{Nuria P. F. Lorente}
\aafil{AAO}

\author{Jeremy Mould\am{Swin}}
\aafil{Swin}

\author{Danail Obreschkow\am{ICRAR}}
\aafil{ICRAR}

\author{Matt S. Owers\am{AAO}}
\aafil{AAO}
\aafil{MQ}

\author{Samuel N. Richards\am{AAO}\am{Syd}\am{CO}}
\aafil{AAO}
\aafil{SIfA}
\aafil{CAASTRO}

\author{Aaron S.G. Robotham}
\affiliation{SUPA School of Physics \& Astronomy, University of St Andrews, KY16 9SS, Scotland}

\author{Adam L. Schaefer\am{Syd}}
\aafil{SIfA}
\aafil{AAO}
\aafil{CAASTRO}

\author{Sarah M. Sweet\am{RSAA}}
\aafil{Swin}

\author{Dan S. Taranu\am{CO}}
\aafil{CAASTRO}
\aafil{ICRAR}

\author{Edoardo Tescari\am{CO}}
\aafil{CAASTRO}
\aafil{UMelb}

\author{Chiara Tonini}
\aafil{UMelb}

\author{T. Zafar\am{AAO}}
\aafil{AAO}

%\begin{document}

\begin{abstract}
  We present the first major release of data from the SAMI Galaxy
  Survey. This data release focuses on the emission-line physics of
  galaxies. Data Release One includes data for 772 galaxies, about
  20\% of the full survey.  Galaxies included have the redshift range
  $0.004 < z < 0.092$, a large mass range ($7.6 < \log \Mstellar/\Msun
  < 11.6$), and star-formation rates of $\sim10^{-4}$ to
  $\sim10^1\Msunyr$. For each galaxy, we include two spectral cubes
  and a set of spatially resolved 2D maps: single- and multi-component
  emission-line fits (with dust extinction corrections for strong
  lines), local dust extinction and star-formation rate. Calibration
  of the fibre throughputs, fluxes and
  differential-atmospheric-refraction has been improved over the Early
  Data Release.
  The data have average spatial resolution of $\n{meanSeeing}$~arcsec
  (FWHM) over the 15~arcsec diameter field of view and spectral
  (kinematic) resolution $R=\n{specR-red}$ ($\sigma=30\kms$) around
  \Ha. The relative flux calibration is better than 5\% and absolute
  flux calibration better than $\pm0.22$~mag, with the latter estimate
  limited by galaxy photometry. The data are presented online through
  the Australian Astronomical Observatory's Data Central.
\end{abstract}
%                              (167 words)


\maketitle

\section{Introduction}

Our textbooks provide a reasonable picture of how the first dark
matter structures assembled out of the primordial matter perturbations
\citep{1999coph.book.....P,2010gfe..book.....M}. But just how gas
settled into these structures to form the first stars and galaxies,
and how these evolved to provide the rich diversity of galaxies we see
around us today, remains an extremely difficult problem to unravel.

Over the past twenty years, imaging surveys from the Hubble Space
Telescope (far field) and the Sloan Digital Sky Survey (near field)
have been particularly effective in identifying evolution of galaxy
parameters with cosmic time and with environment across large-scale
structure. This has been matched by extensive surveys using
multi-object spectroscopy
\citep[e.g.][]{2000AJ....120.1579Y,2001MNRAS.328.1039C,2011MNRAS.413..971D}
that have usually provided a single spectrum within a fixed fibre
aperture at the centre of each galaxy; spatial information must be
drawn from multi-wavelength broadband images.

It has long been recognized that large-scale multi-object
spectroscopic surveys do not provide a complete picture of
galaxies. The complexity of galaxies cannot be captured with a single
average or central spectrum. Three-dimensional imaging spectroscopy,
or \ac{IFS} is needed to quantify each galaxy.

Driven by pioneering work using Fabry-Perot interferometry
\citep{1974ApJS...27..415T} and lenslet arrays
\citep{1988igbo.conf..266C}, \ac{IFS} has exploited the plunging costs
of large-area detectors to dominate extra-galactic studies today
\citep[e.g.][]{2014AdOT....3..265H}.  The first generation of IFS
surveys, sampling 10s to 100s of galaxies, have only recently
completed.  Examples include ATLAS${}^\textrm{3D}$
\citep{2011MNRAS.413..813C}, CALIFA \citep{2012A&A...538A...8S} and
SINS \citep{2009ApJ...706.1364F}.  These surveys demonstrated that
there is much to learn from both the stellar and gaseous components in
data of this kind.  However, these surveys all used instruments that
target individual galaxies one at a time and are, therefore, not
optimal for surveying thousands of galaxies.  To move beyond
catalogues of a few hundred requires effective multiplexing.
%                              (131 words)

Multiplexing IFS has only recently become possible.  The FLAMES
instrument on the VLT \citep{2002Msngr.110....1P} was the first, with
15 integral-field units (IFUs) each having 20 spatial resolution
elements in a $2\times3$ arcsec field of view.

The two main approaches to \ac{IFS} are fibre-based and slicer-based
systems.  Slicers have higher sensitivity below 400 nm and in the
infrared as shown by the KMOS instrument on the VLT
\citep{2013Msngr.151...21S}; they also have excellent performance over
narrow fields of view, particularly when assisted by adaptive optics
\citep[NIFS;][]{2003SPIE.4841.1581M}.  However, fibres ease deployment
of IFUs over wide fields of view and allow the spectrograph to be
mounted on the floor rather than on the telescope, simplifying design
and improving stability.  Fibre based systems are therefore preferred
for wide-field, multi-object \ac{IFS} in the optical bands.

With the aim of carrying out \ac{IFS} surveys targeting thousands of
galaxies, we developed the Sydney/AAO Multi-object Integral-field
spectrograph \citep[SAMI;][]{2012MNRAS.421..872C} on the 3.9m
Anglo-Australian Telescope.  SAMI provides a multiplex of $\times 13$
with each \ac{IFU} having a diameter of 15 arcsec and uses compact
fused fibre bundles with minimised cladding between the fibre cores
\citep[hexabundles:][]{2011OpExpr.19.2649,
  2011MNRAS.415.2173,2014MNRAS.438.869}.  The MaNGA Survey
\citep{2015ApJ...798....7B} operating on the Apache Point 2.3m
Telescope, has also begun a similar project, with an \ac{IFU}
multiplex of $\times 16$.  Meanwhile, the high-redshift KMOS-3D and
KROSS Surveys \citep{2015ApJ...799..209W, 2016MNRAS.456.4533M} are
making spatially resolved observations of high redshift galaxies.

Large-scale IFS surveys are uniquely positioned to address a number of
the outstanding questions regarding galaxy formation and evolution
\citep{2012MNRAS.421..872C,2015IAUS..309...21B}, including:
\begin{itemize}
\item What is the physical role of environment in galaxy evolution?
\item What is the interplay between gas flows and galaxy evolution?
\item How are mass and angular momentum built up in galaxies?
\end{itemize}

Mass is thought to be the primary discriminant driving the huge
variety of galaxies observed, setting their star formation rate
\citep[e.g.][]{2010ApJ...721..193P, 2012ApJ...757....4P}, metallicity
\citep[e.g.][]{2004ApJ...613..898T}, and morphology.  However, in
addition to mass, the environment of a galaxy also plays a central
role in controlling such properties (e.g.\
\citealt{2002MNRAS.334..673L,2009ARA&A..47..159B}, and
\citealt{1980ApJ...236..351D,2011MNRAS.416.1680C},
respectively). Despite the wealth of data at hand, the physical
processes that drive environmental differences are still
uncertain. The processes are likely to depend on whether a galaxy is
the central galaxy or a satellite in its parent halo, the mass of the
parent halo, and local galaxy--galaxy interactions
\citep[e.g.][]{2015MNRAS.452..616D}. With the broad range of
observables available to SAMI, we can directly test which physical
processes are at play in environmental transformations.

Gas flow (or lack thereof) in and out of a galaxy controls its
evolution with time. Inflows have formed disks, fuelled generation
upon generation of new stars, and fed supermassive black holes. In
current galaxy-formation theory, galactic-scale outflows explain the
difference between the theoretical cold-dark-matter mass function and
the observed stellar-mass function
\citep[e.g.][]{2012MNRAS.421..621B}. A feedback process with strong
mass dependence is needed to resolve this problem.  Outflows offer the
most promising solution \citep[e.g.][]{2012RAA....12..917S}, and are
clearly detected by combining gaseous emission-line ionisation
diagnostics with kinematics \citep[e.g.][]{2010ApJ...711..818S,
  2012ApJ...761..169F, 2014MNRAS.444.3894H, 2016MNRAS.457.1257H}.  Gas
inflows can be traced using the measurement of misalignment between
gas and stellar kinematics
\citep[e.g.][]{2011MNRAS.417..882D,2016MNRAS.457..272D} and by
searching for flattened metallicity gradients
\citep{2010ApJ...721L..48K,2012ApJ...753....5R}.

The mass and angular momentum of a galaxy are most directly probed by
its kinematic state. A galaxy's accretion and merger history is
central to defining its character, and aspects of this history are
encoded in the line-of-sight velocity distributions. By studying the
detailed kinematics of galaxies across the mass and environment plane,
we unlock a new view of galaxy evolution
\citep{2016MNRAS.463..170C,2017ApJ...835..104V}.  IFS has defined a
new set of morphological classifications in terms of dynamical
properties \citep[e.g.][]{2011MNRAS.414..888E, 2011MNRAS.413..813C},
such as the separation into fast rotators (rotation dominated) and
slow rotators (dispersion dominated).  We aim to understand how these
kinematic properties are distributed across the mass--environment
plane, and to make direct comparison to simulations that are now
becoming available to measure more complex dynamical signatures
\citep[e.g.][]{2014MNRAS.444.3357N}.

IFS surveys have arrived at an auspicious time.  Cosmological-scale
hydrodynamic simulations can now form thousands of galaxies with
realistic properties in $\sim100\Mpc^3$ volumes
\citep[e.g.][]{2014MNRAS.444.1518V,2015MNRAS.446..521S}.  These
simulations allow study of how gas enters galaxies
\citep[e.g.][]{2012MNRAS.427.3320C} and the impact of feedback
\citep[e.g.][]{2015ApJ...804L..40G}.  Those at higher resolution
\citep[e.g.][]{2009ApJ...694..396B,2014MNRAS.445..581H} are probing
details of disk formation, gas flows and feedback, though not yet
within a full cosmological context.  Direct, detailed comparison of
spatially resolved data to these simulations is required to advance
our understanding.

In this paper, we present Data Release One (DR1) of the SAMI Galaxy
Survey, building on our Early Data Release (EDR) in 2014
\citep[see][]{2015MNRAS.446.1567A}.  We provide data cubes for
\n{nsample} galaxies and value-added products based on detailed
emission-line fitting.  Future releases will provide more galaxies and
products.  In Section \ref{sec:samigalaxysurvey} we review the SAMI
Galaxy Survey itself, including the selection, observations, data
reduction and analysis.  In Section~\ref{sec:data-products} we
describe the Core data being released, with discussion of data quality
in subsection~\ref{sec:quality}. The emission-line-physics value-added
products are described in Section~\ref{sec:vap}. The online database
is introduced in Section~\ref{sec:database}. We summarise in Section
\ref{sec:future}.  Where required, we assume a cosmology with
$\Omega_m=0.3$, $\Omega_\Lambda=0.7$ and $H_0=70\kms\Mpc^{-1}$.

\section{Brief Review of the SAMI Galaxy Survey}
\label{sec:samigalaxysurvey}

The SAMI Galaxy Survey is the first integral-field spectroscopic
survey of enough galaxies to characterise the spatially-resolved
variation in galaxy properties as a function of both mass and
environment.  Specific details concerning the survey can be found in
papers describing the SAMI instrument
\citep{2012MNRAS.421..872C,2015MNRAS.447.2857B}, the SAMI-GAMA Sample
Target Selection \citep{2015MNRAS.447.2857B}, the SAMI Cluster Sample
Target Selection \citep{2017arXiv170300997O}, data reduction
\citep{2015MNRAS.446.1551S} and the Early Data Release
\citep{2015MNRAS.446.1567A}.  Below we review key aspects of the
survey.

\subsection{The SAMI instrument}
\label{sec:inst}

SAMI is mounted at the prime focus of the Anglo-Australian Telescope
and has 1-degree-diameter field of view. SAMI uses 13 fused optical
fibre bundles \citep[hexabundles;][]{2011OpExpr.19.2649,
  2011MNRAS.415.2173, 2014MNRAS.438.869} with a high (75 percent) fill
factor. Each bundle combines 61 optical fibres of 1.6 arcsec diameter
to form an \ac{IFU} of 15-arcsec diameter. The 13 \acp{IFU} and 26 sky
fibres are inserted into pre-drilled plates using magnetic
connectors. Optical fibres from SAMI feed into AAOmega, a
bench-mounted double-beam optical spectrograph
\citep{2006SPIE.6269E..14S}. AAOmega provides a selection of different
spectral resolutions and wavelength ranges. For the SAMI Galaxy
Survey, we use the 580V grating at $3700-5700$\AA\ and the 1000R
grating at $6250-7350$\AA. With this setup, SAMI delivers a spectral
resolution of $R=\n{specR-blue}$ ($\sigma=\n{specDeltaSigma-blue}$)
for the blue arm, and $R=\n{specR-red}$
($\sigma=\n{specDeltaSigma-red}$) for the red arm at their respective
central wavelengths \citep{2017ApJ...835..104V}. A dichroic splits the
light between the two arms of the spectrograph at 5700\AA.
%                              (165 words)

\subsection{Target Selection}
\label{sec:target-sel}

In order to cover a large dynamic range in galaxy environment, the
SAMI Galaxy Survey is drawn from two regions with carefully matched
selection criteria.  The majority of targets are from the Galaxy And
Mass Assembly (GAMA) Survey \citep{2011MNRAS.413..971D}, and we denote
this as the SAMI-GAMA Sample.  However, the volume of the SAMI-GAMA
region does not contain any massive galaxy clusters, so a second set
of targets are drawn from specific cluster fields.  This we denote as
the SAMI Cluster Sample \citep{2017arXiv170300997O}.

DR1 includes galaxies only from the SAMI-GAMA Sample and the selection
for these targets is described by \cite{2015MNRAS.447.2857B}. Briefly,
the sample is drawn from the $4\times12$-degree fields of the initial
GAMA-I survey \citep{2011MNRAS.413..971D}, but uses the deeper
spectroscopy to $r<19.8$ of the GAMA-II sample
\citep{2015MNRAS.452.2087L}.  The high completeness of the GAMA sample
(98.5 per cent) leads to high-reliability group catalogues
\citep{2011MNRAS.416.2640R} and environmental metrics
\citep{2013MNRAS.435.2903B}.  The GAMA regions also provide broad-band
imaging from the ultraviolet to far-infrared
\citep{2016MNRAS.455.3911D}.

\begin{figure*}
  \centering  
  \includegraphics[width=1\linewidth]{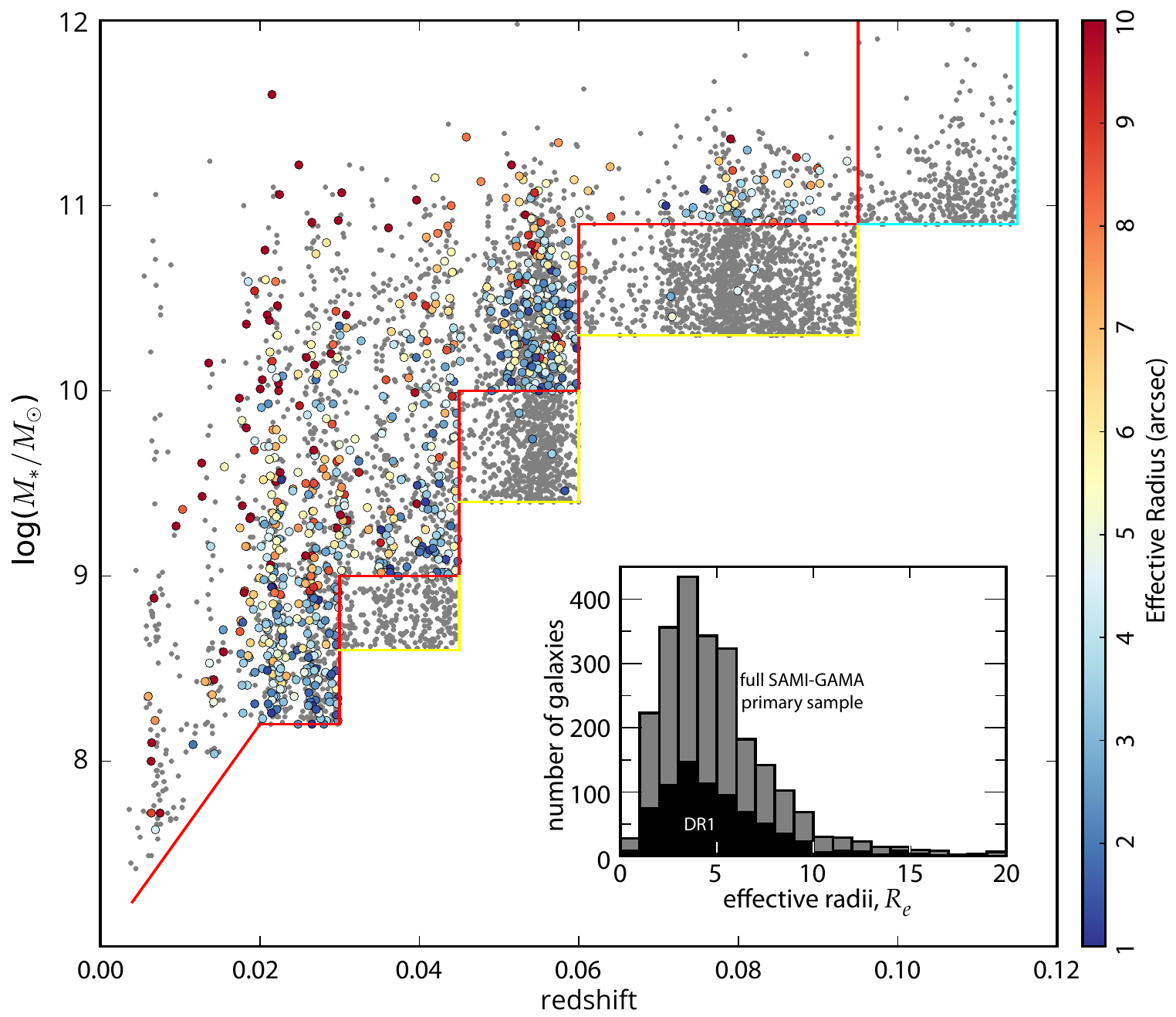}
  \caption{%
    \label{fig:z_Mstar_Re}%
    The SAMI-GAMA portion of Galaxy Survey targets in the redshift
    vs.\ stellar mass plane. The primary targets lie above the red
    line, and secondary targets lie above the cyan (higher
    redshift) or yellow (lower mass) line. Light grey points show the
    full SAMI-GAMA sample, while the targets comprising DR1 are
    coloured by effective radius ($R_e$) in arcsec. The inset
    histogram illustrates that the $R_e$ distribution of the DR1
    galaxies (black) is representative of the full primary sample
    (grey). %
  }
\end{figure*}

The selection limits for the SAMI-GAMA Sample, shown in
Figure~\ref{fig:z_Mstar_Re}, consist of a set of volume-limited
samples with stellar-mass limits stepped with redshift. We select
using stellar masses determined from only \emph{g}- and \emph{i}-band
photometry and redshift, using the relationship given in Eq.~3 of
\cite{2015MNRAS.447.2857B}.  This determination is based on the
relationship between mass-to-light ratio and colour derived by
\cite{2011MNRAS.418.1587T}, and assumes a \cite{2003PASP..115..763C}
initial-mass function.
%                              (170 words)

\subsection{Observing strategy}
\label{sec:obs-strategy}

\cite{2015MNRAS.447.2857B} describe the process of allocating target
galaxies to fields for observation.

Our standard observing sequence consists of a flat-field frame (from
the illuminated AAT dome) and arc frame, followed by seven object
frames each of 1800~s exposure.  A flat field and arc are taken to end
the sequence.  The seven object exposures are offset from one another
in a hexagonal dither pattern \cite[see][Fig.16]{2015MNRAS.447.2857B},
with the subsequent frames radially offset from the first exposure by
$0.7"$ in each of six directions 60 degrees apart.  This offset is
applied based on the most central guide star in the field, using an
offset in pixels on the guide camera.  Variations in atmospheric
refraction and dispersion between different exposures causes the
effective offsets to differ for different galaxies on the same field
plate.  However, the high fill factor of SAMI hexabundles minimises
the effect on data quality (see especially
Section~\ref{sec:subseeing}).  The change in offset across the field
is measured as part of the alignment process during data reduction as
described in \cite{2015MNRAS.446.1551S}.

Where possible, twilight-sky frames are taken for each field to
calibrate fibre-throughput. Primary spectrophotometric standards are
observed each night that had photometric conditions to provide
relative flux calibration (i.e.\ the relative colour response of the
system).

\subsection{Data reduction}
\label{sec:dr}

Raw telescope data is reduced to construct spectral cubes and other
core data products in two stages that are automated for batch
processing using the ``SAMI Manager'', part of the \texttt{sami}
python package \citep{2014ascl.soft07006A}.  The specifics of both
stages are detailed in \cite{2015MNRAS.446.1551S}.  Subsequent changes
and improvements to the process are described in section~3 of
\cite{2015MNRAS.446.1567A}\ and in Section~\ref{sec:dr-changes} below.

The first stage of data reduction takes raw 2D detector images to
partially calibrated spectra from each fibre of the instrument,
including spectral extraction, flat-fielding, wavelength calibration
and sky subtraction.  Processing for this stage uses the \emph{2dfdr}
fibre data reduction package \citep{2015ascl.soft05015A} provided by
the Australian Astronomical Observatory\footnote{Different versions of
  \emph{2dfdr} are available, along with the source code for more
  recent versions at
  \url{http://wwww.aao.gov.au/science/software/2dfdr}}. This stage
outputs the individual fibre spectra as an array indexed by fibre
number and wavelength, and referred to as ``row-stacked spectra'' (RSS).
%                              (108 words)

In the second stage, the row-stacked spectra are sampled on a regular
spatial grid to construct a 3-dimensional (2 spatial and 1 spectral)
cube. Processing for the second stage is done within the \texttt{sami}
python package \citep{2014ascl.soft07006A}. This stage includes
telluric correction, flux calibration, dither registration,
differential atmospheric refraction correction and mapping input
spectra onto the output spectral cube.  The last of these stages uses
a drizzle-like algorithm \citep{2002PASP..114..144F,
  2015MNRAS.446.1551S}.  The spectral cubes simplify most subsequent analysis
because the cube can be read easily into various packages and
programming languages, and spatial mapping of the data is
straightforward. However, in creating the spectral cube, additional
covariance between spatial pixels is introduced that must be correctly
considered when fitting models and calculating errors
\citep{2015MNRAS.446.1551S}.

\subsection{Comparing SAMI with other large IFS Surveys}
\label{sec:sami-vs-others}

\paragraph{Spatial resolution} %
The SAMI Galaxy Survey has less spatial resolution elements per galaxy
than most first generation IFS surveys. First generation surveys were
based on instruments with a single IFU with a large field of view on
the sky and many spatial samples.  For example, CALIFA uses the PPAK
fibre bundle \citep{2006PASP..118..129K} that contains 331 science
fibres and uses this bundle to target a single galaxy at a time. In
contrast, SAMI has 793 target fibres, a factor of $\times2.4$ more,
but distributes them over 13 targets, with a much smaller field of
view per IFU.  The ATLAS${}^\textrm{3D}$ and CALIFA Surveys target
lower redshift galaxies better matched in size to their larger IFUs,
leading to higher spatial resolution. Therefore, these first
generation surveys continue to serve as a benchmark for local
($<100\Mpc$) galaxies, while second generation surveys will provide
much larger samples of slightly more distant galaxies (typically
$>100\Mpc$).
%                              (150 words)

\paragraph{Spectral resolution} %
In the neighbourhood of the \Ha\ emission line, the SAMI Galaxy Survey
has higher spectral resolution than most other first- and
second-generation surveys.  In the blue arm the large number of
spectral features visible drives the survey design to broad wavelength
coverage (3700--5700\AA), leading to a resolution of $R\simeq
\n{specR-blue}$.  However, in the red arm, by limiting spectral
coverage to a $\sim1100$\AA\ region around the \Ha\ emission line we
can select a higher spectral resolution, $R\simeq\n{specR-red}$.  This
selection is distinct from most other surveys, such as CALIFA and
MaNGA, with $R\simeq850$ and $R\simeq2000$ respectively around the
\Ha\ line. Therefore, analyses based on SAMI data can better separate
distinct kinematic components \citep[e.g.\ in outflows;
see][]{2014MNRAS.444.3894H,2016MNRAS.457.1257H}, can more accurately
measure the gas velocity dispersion in galaxy disks
\citep{FederrathEtAl2017}, and can investigate the kinematics of dwarf
galaxies.  The trade-off for the higher spectral resolution in the red
arm is more limited spectral coverage, that only extends to
$\sim7400$\AA, whereas MaNGA reaches to $\sim1$\,$\mu$m.
%                              (189 words)

\paragraph{Environment measures} %
The SAMI Galaxy Survey also benefits from more complete and accurate
environmental density metrics than other IFS surveys.  The GAMA Survey
has much greater depth ($r<19.8$ vs $r<17.8$) and spectroscopic
completeness ($>98$ per cent vs $\simeq 94$) than the SDSS on which
the MaNGA Survey is based (\citealt{2011MNRAS.413..971D} and
\citealt{2015ApJS..219...12A}, respectively). Therefore, GAMA provides
several improved environmental metrics over SDSS, including group
catalogues and local-density estimates (\citealt{2011MNRAS.416.2640R}
and \citealt{2013MNRAS.435.2903B}, respectively).  For example, 58 per
cent of primary Survey targets are members of a group identified from
GAMA \citep[containing two or more galaxies based on a
friends-of-friends approach--see][]{2011MNRAS.416.2640R}, but only 15
per cent are members of a group identified from SDSS
\citep{2007ApJ...671..153Y}.
%                              (123 words)

\paragraph{Range in mass} %
The SAMI Survey provides a broader range in mass of galaxies than
MaNGA at the expense of more variability in the radial coverage of
galaxies. Our target selection aims to be 90 percent complete above
the stellar-mass limit for each redshift interval targeted while
covering a large range in stellar mass ($8 \lesssim \log(M_* / \Msun)
\lesssim 11.5$).  This selection results in a more extensive sampling
of low-mass galaxies than previous surveys. It also differs from the
MaNGA selection, which targets galaxies in a relatively narrow
luminosity range at each redshift. The MaNGA selection leads to less
variability in the radial extent of the data relative to galaxy size.

\paragraph{Sampling of galaxy clusters} % 
The Survey's cluster sample is also unique among IFS surveys.  Massive
clusters are rare, so volume-limited samples typically include few
galaxies belonging to these extreme environments. However, only in
clusters are the extremes of environmental effects demonstrated on
galaxy evolution. With the Survey's cluster sample, one can trace in
detail the evolution of galaxies in the densest environments.  Other
programs have targeted individual clusters for IFS observations
\citep[e.g.][]{2011MNRAS.413..813C,2013MNRAS.436...19H,2013MNRAS.429.1258D,2014MNRAS.441..274S},
but the SAMI cluster sample is the most comprehensive IFS study of
clusters yet attempted.  The SAMI Galaxy Survey sample includes eight
different clusters (APMCC0917, A168, A4038, EDCC442, A3880, A2399,
A119 and A85), allowing investigation of variability between
clusters. Part of the Survey includes new (single-fibre) multi-object
spectroscopy of these clusters to ascertain cluster membership, mass,
and dynamical properties \citep{2017arXiv170300997O}.

%                              (71 words)

\section{Core data release}
\label{sec:data-products}

The galaxies included in DR1 are drawn exclusively from the SAMI-GAMA
Sample. The included core data products are the regularly gridded flux
cubes (spectral cubes). All of the core data included have met minimum
quality standards, and the quality of the final data has been measured
with care.

\subsection{Galaxies included in DR1}
\label{sec:dr1-sample}

\begin{figure*}
  \centering  
  \includegraphics[width=0.95\linewidth]{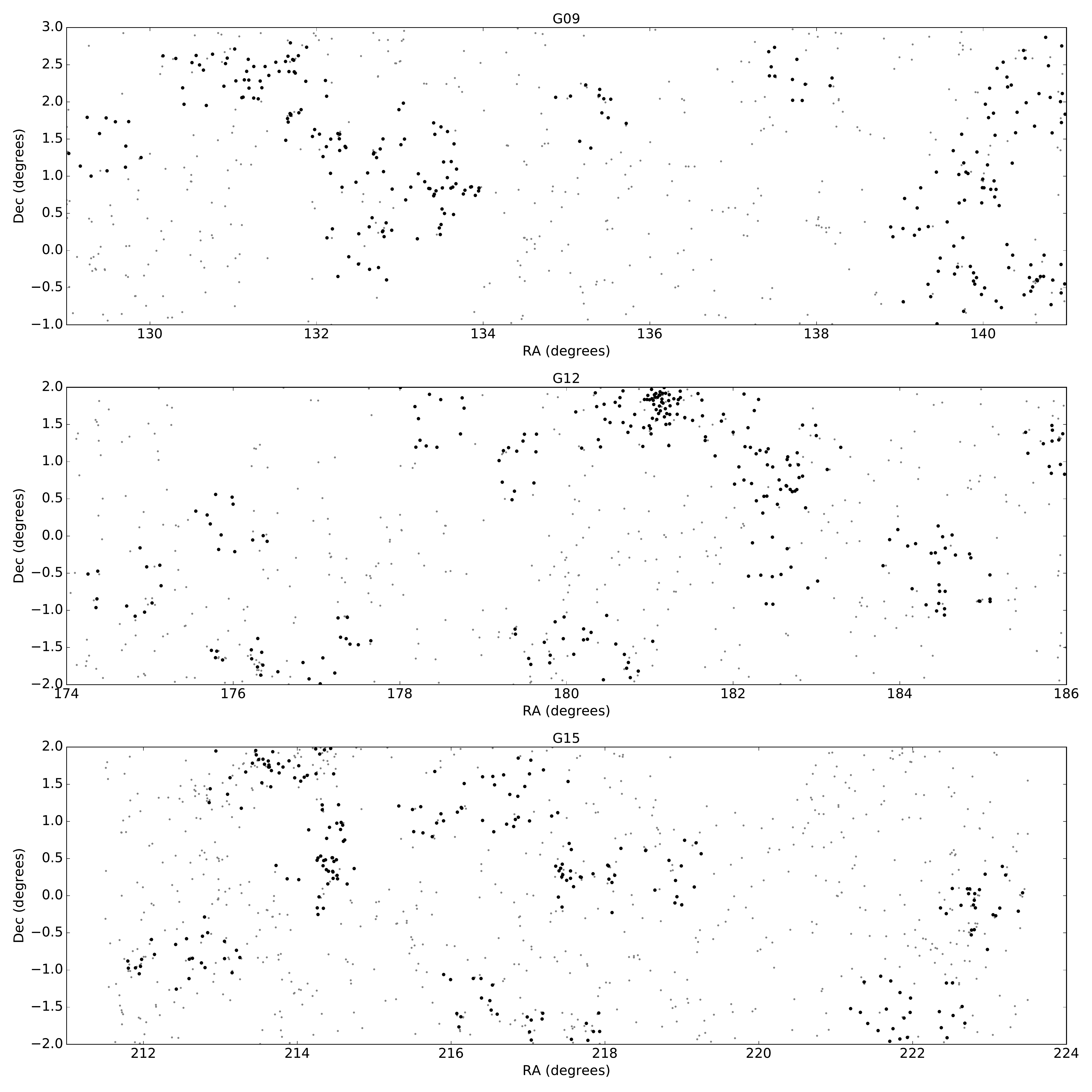}  
  \caption{%
    \label{fig:ra_dec}%
    Distribution on the sky of the SAMI-GAMA Sample, covering GAMA regions
    G09, G12, and G15. The primary targets of the complete field sample
    are shown by the small grey points, targets included in this DR1
    are shown in black. %
  }
\end{figure*}

\begin{figure*}
  \centering  
  \includegraphics[width=1\linewidth]{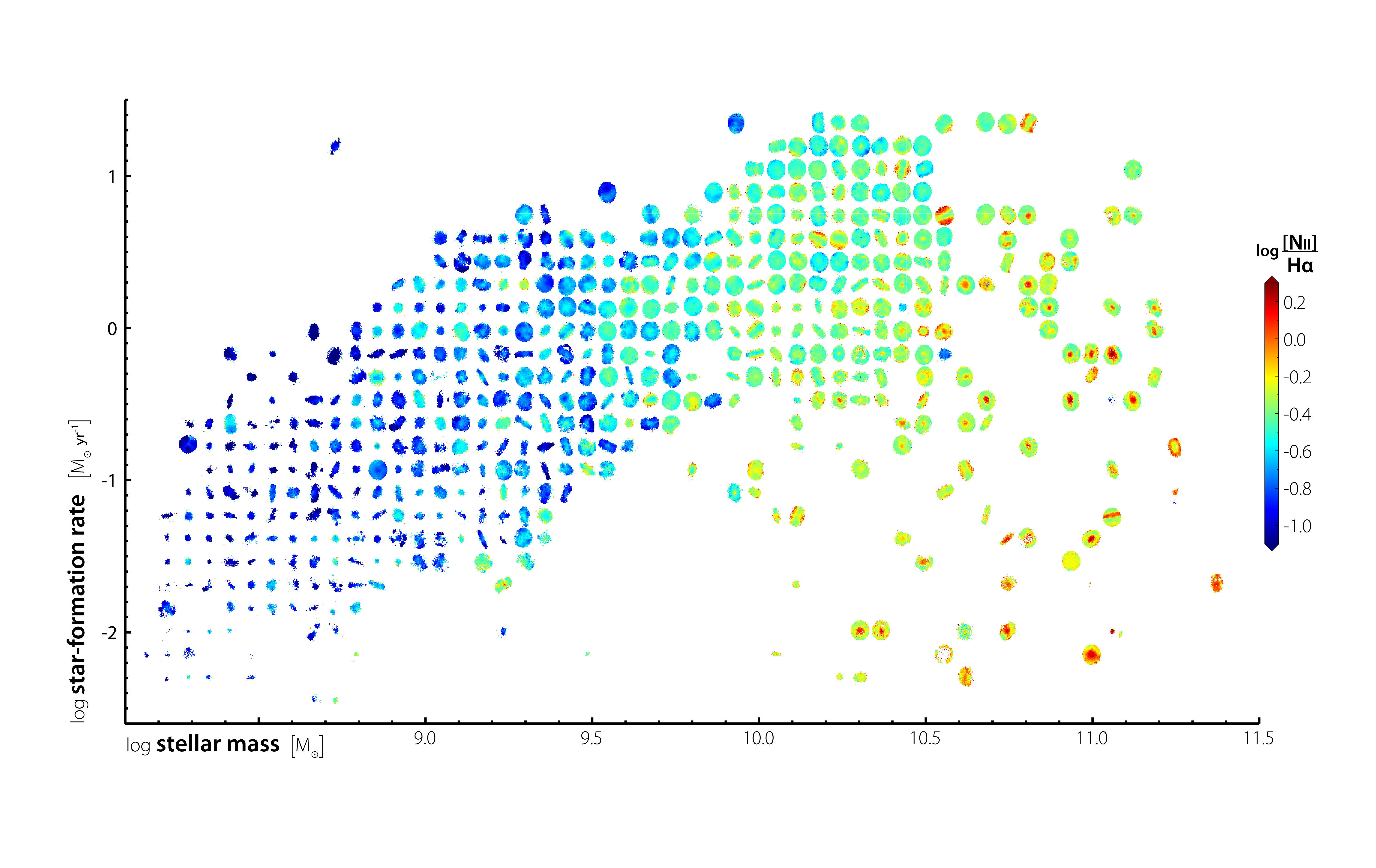}
  \caption{%
    \label{fig:poster-plot}%
    The spatially-resolved maps of {\sc [Nii]}/\Ha{} within 15 arcsec
    diameter of DR1 galaxies are arrayed by stellar mass and
    star-formation rate. Not all in DR1 appear because some have
    insufficient {\sc [Nii]} and/or \Ha{} flux for their S/N ratio to
    exceed 3 across their extent.  Some maps have been shifted
    slightly to avoid overlap, so stellar masses and star-formation
    rates shown are indicative, not
    exact. %
  }
\end{figure*}

Galaxies in DR1 are drawn from all \n{nsample0} galaxies observed in
the SAMI-GAMA sample through June, 2015 (AAT semesters 2013A to
2015A). This includes all galaxies in the Survey's EDR (but the data
for those galaxies have been reprocessed for this
Release). Table~\ref{tab:Obs} shows how the DR1 galaxy numbers compare
to the current progress of the SAMI Galaxy Survey in the GAMA
regions. The distribution of these targets in the stellar
mass--redshift plane, on the sky and in the star formation
rate--stellar mass plane can be seen in Figures \ref{fig:z_Mstar_Re},
\ref{fig:ra_dec} and \ref{fig:poster-plot}, respectively.

\begin{table}
  \centering
  \caption{% 
    \label{tab:Obs} %
    SAMI-GAMA Sample primary and filler targets (see 
    Figure~\ref{fig:z_Mstar_Re}) observed by end of 2016 and their DR1 
    release. %
  }
  \begin{tabular}{l c c c}
    \hline\hline
    & No. in  & Galaxies  & Galaxies in  \\ 
    & catalogue &  observed & this Release  \\

    \hline
    Primary targets   & 2404    &    1267          &  763 \\
    Filler targets   & 2513     &    44             &     9 \\
    \hline
  \end{tabular}
\end{table}

We have not included some observed galaxies in DR1 for quality control
reasons.  From the \n{nsample0} galaxies, we removed those with:
\begin{itemize}
\item fewer than 6 individual exposures meeting the minimum standard
  of transmission greater than 0.65 and seeing less than 3~arcsec FWHM
  (48 galaxies removed); and
\item individual observations that span more than one month for a
  single field and have differences in their heliocentric velocity
  frames of greater than 10 km/s (12 galaxies removed).
\end{itemize}
After removing observations that did not meet these data quality
requirements, \n{nsample} galaxies remain. 

Galaxies included in DR1 may have a small bias towards denser
regions over the full field sample. The order in which galaxies are
observed over the course of the Survey is set by the tiling process,
which allocates galaxies to individual observing fields. Tiling is
based only on the sky distribution of galaxies---not their individual
properties.  Initial tiles are allocated preferentially to regions
with higher sky density to maximize the efficiency of the Survey over
all. Figure~\ref{fig:ra_dec} shows the three GAMA-I fields (G09, G12 and
G15) and the sky distribution of galaxies in this data release
compared with the overall SAMI field sample.
%                              (105 words)

DR1 galaxies are distributed across the full range of the primary
sample in redshift, stellar mass and effective radius as illustrated
in Figure~\ref{fig:z_Mstar_Re}. A Kolmogorov--Smirnov test indicates
that the DR1 sample has the same effective radius distribution as the
SAMI field sample (D-statistic=0.025, p-value=0.85). However there is
a difference in the distribution of stellar mass (D-statistic=0.08,
p-value=0.001), such that lower mass galaxies are slightly over
represented in the DR1 sample.

\subsection{Changes in data reduction methods since the Early Data Release}
\label{sec:dr-changes}

For DR1 we use the \texttt{sami} python package snapshot identified
as Mercurial changeset \texttt{0783567f1730}, and \emph{2dfdr} version
5.62 with custom modifications. The version of \emph{2dfdr} is the
same as for our Early Data Release \citep{2015MNRAS.446.1567A}, and
all of the modifications are described by
\cite{2015MNRAS.446.1551S}.
These changes have been integrated into subsequent public release
versions of \emph{2dfdr}. Changes in the \texttt{sami} package are
described in the rest of this section.

\subsubsection{Fibre throughput calibration}
\label{sec:fibre-thro-calibr}

To achieve good flux calibration and uniform image quality, the
relative throughput of each of the \n{fibres} fibres (including 26 sky
fibres) must be normalised to a common value. We have improved the
approach for normalising the fibre throughputs over that used in our
EDR.

The fibre-throughput calibration used in our EDR had two shortcomings
that limited data quality, particularly from the blue arm of the
spectrograph.  In our EDR, the relative throughput of individual
fibres was primarily determined from the integrated flux in the
night-sky lines for long exposures, and from the twilight flat-fields
for short exposures.  However, the blue data (3700--5700\AA) include
only one strong night sky line, 5577\AA, so are particularly
susceptible to two problems.  First, sky lines are occasionally
impacted by cosmic rays, leading to poor throughput estimates for
individual fibres.  Second, the limited photon counts in the sky line
limits the estimates of the relative throughput to $\simeq 1-2$ per
cent.
 
For DR1 the relative fibre throughputs were calibrated from either
twilight flat-field frames, or from dome flat-field frames for fields
where no twilight flat was available. The night sky spectrum was then
subtracted using this calibration. If the residual flux in sky spectra
was excessive (mean fractional residuals exceeded 0.025), then the
fibre throughputs were remeasured using the integrated flux in the
night-sky lines (as in the EDR). If all sky lines in a fibre were
affected by bad pixels (typically only an issue for the blue
wavelength range, which covers only a single sky line), then the mean
fibre-throughput calibration derived from all other frames of the same
field was adopted. The sky subtraction was then repeated with the
revised throughput values. The method that provided the final
throughput calibration is listed with the cubes in the online
database. This approach ensures that, for the calibration options
available, the best option is used to calibrate the fibre throughputs.

\subsubsection{Flux calibration}
\label{sec:flux-cal}

The flux calibration process has been improved over our EDR to better
account for transparency changes between individual observations of a
field and improve overall flux calibration accuracy.  In our EDR, the
absolute flux calibration was applied after forming all cubes for a
field of 12 galaxies and 1 secondary standard star. All objects in the
field were scaled by the ratio of the field's secondary standard star
observed {\it g-}band flux to the SDSS photometry \emph{after
  combining individual observations into cubes} \cite[for full details
see section~4.4 of][]{2015MNRAS.446.1551S}.

For DR1 this scaling has also been applied to each individual RSS
frame for a given field \emph{before forming cubes}, i.e., the scaling
is now applied twice. This additional scaling ensures that differences
in transparency between individual observations are removed before the
cube is formed, which improves the local flux calibration accuracy and
removes spatial `patchiness' in the data. The accuracy of the overall
flux calibration is discussed in Section~\ref{sec:dq-flux-cal}.
%                              (71 words)

\subsubsection{Differential atmospheric refraction correction}
\label{sec:dar-corr}

For DR1 we have improved the correction for differential
atmospheric refraction over that in our EDR. The
atmospheric dispersion is corrected by recomputing the drizzle
locations of the cube at regular wavelength intervals \cite[see
section~5.3 of][]{2015MNRAS.446.1551S}. In our EDR the
drizzle locations were recomputed when the accumulated dispersion
misalignment reached 1/10th of a spaxel (0.05~arcsec). We found
that this frequency caused unphysical `steps' in the spectra
within a spaxel. In DR1 we recalculated the drizzle
locations when the accumulated dispersion misalignment reached 1/50th
of a spaxel, i.e., five times more often than in the Early Data
Release. This significantly reduced the impact of atmospheric
dispersion on the local flux calibration within individual
spaxels. Section~\ref{sec:subseeing} elaborates on how atmospheric
dispersion affects the quality of the data.
%                              (134 words)

\subsection{Core Data Products included}
\label{sec:core-data}

Several Core data products are included in DR1: flux spectral
cubes with supporting information, GAMA catalogue data used for the
target selection, and Milky Way extinction spectra.

\subsubsection{Spectral Cubes}
\label{sec:spectral-cubes}

The position--velocity spectral flux cubes are the
products most users will value. These cubes are presented with
the following supporting data, all sampled on the same regular grid:
\begin{description}
\item[variance] The uncertainty of the intensities as a variance,
  including detector-readout noise and Poisson-sampling noise
  propagated from the raw data frames.
\item[spatial covariance] co-variance between adjacent spatial pixels
  introduced by drizzle mapping onto the regular grid. The
  co-variance and the format of this five-dimensional array are
  described in section~5.7 of \cite{2015MNRAS.446.1551S}.
\item[weights] The effective fractional exposure time of each pixel,
  accounting for gaps between individual fibres, dithering, etc. These
  are described in section~5.3 of
  \citet{2015MNRAS.446.1551S}.
\end{description}

A world-coordinate system (WCS) for each cube is
included. This WCS maps the regular grid onto sky- (right
ascension and declination) and wavelength-coordinates. The origin of
the spatial coordinates in the WCS is defined using a 2D Gaussian fit
to the emission in the first frame of the observed dither sequence. The
wavelength coordinates are defined in the data-reduction process from
arc-lamp frames. The accuracy of the spatial coordinates is discussed
in Section~\ref{sec:wcs-accuracy} and that of the wavelength coordinate in
section~5.1.3 of \citet{2015MNRAS.446.1567A}.

Also provided for each spectral cube are estimates of the point-spread
function (PSF) of the data in the spatial directions. The PSF is
measured simultaneously with data collection using the
secondary standard star included in each SAMI field. We provide the
parameters of a circular-Moffat-profile fit to that
star image (i.e.\ the flux calibrated red and blue star cubes summed
over the wavelength axis). The Moffat profile has form
\begin{align}
f = \frac{\beta - 1}{\pi \alpha^2}  \left(1 + {\left(\frac{r}{\alpha}\right)}^2\right)^{-\beta}. 
\end{align}
where $\alpha$ and $\beta$ parameterize the fit and $r^2 = x^2 + y^2$
is the free variable denoting spatial position
\citep{1969A&A.....3..455M}. The reported PSF is the luminosity
weighted average over the full (i.e. red + blue) SAMI wavelength
range. With the parameters of the Moffat-profile fit, we also provide
the corresponding FWHM, $W$, as given by
\begin{align}
W = 2 \alpha \sqrt{2^{(1/\beta)} - 1},
\end{align}
measured in arcseconds. The distribution of measured PSF is
discussed in section~5.3.2 of \citet{2015MNRAS.446.1567A}, and
is unchanged in DR1. 

Finally, for convenience, we include the exact versions of the GAMA
data used in the sample selection of the SAMI field sample. Note that
in some cases, newer versions of these data are available from the
GAMA Survey and should be used for scientific analysis.

\subsubsection{Milky Way dust-extinction correction}
\label{sec:mwdust}

SAMI spectral cubes are not corrected for dust extinction, either
internal to the observed galaxy or externally from Milky Way
dust. However, we do provide a dust-extinction-correction curve for
each galaxy to correct for the latter.  Using the right ascension and
declination of a galaxy, we determined the interstellar reddening,
E(B-V), from the Planck v1.2 reddening maps
\citep{2014A&A...571A..11P} and the \cite{1989ApJ...345..245C}
extinction law to provide a single dust-correction curve for each
spectral cube. Note that this curve \emph{has not been applied to the
  spectral cubes}. To correct a SAMI cube for the effects of Milky Way
dust, the spectrum of each spaxel must be multiplied by the
dust-correction curve.

\subsection{Data Quality}
\label{sec:quality}

We now discuss data quality measurements for the Core data released.
\cite{2015MNRAS.446.1567A} discusses the quality of the data in our
EDR, including fibre cross-talk, wavelength calibration, flat-fielding
accuracy, and other metrics. Where data quality does not differ
between our EDR and DR1, we have not repeated the discussion of
\cite{2015MNRAS.446.1567A}.  Instead, we discuss the data-quality
metrics potentially affected by changes in the data reduction.

\subsubsection{Sky Subtraction Accuracy}
\label{sec:continuum-sky}

The changes to fibre throughput calibration (see Section~\ref{sec:dr})
removes occasional (less than one fibre per frame) catastrophically
bad throughputs.  It does not change the overall average sky
subtraction accuracy, as presented by \citet{2015MNRAS.446.1567A}.
The lack of change in sky subtraction precision suggests that fibre
throughput and photon counting noise in the blue 5577\AA\ line is not
currently a limiting factor in the precision of sky subtraction.

\begin{figure*}
  \centering  
  \includegraphics[width=0.5\linewidth]{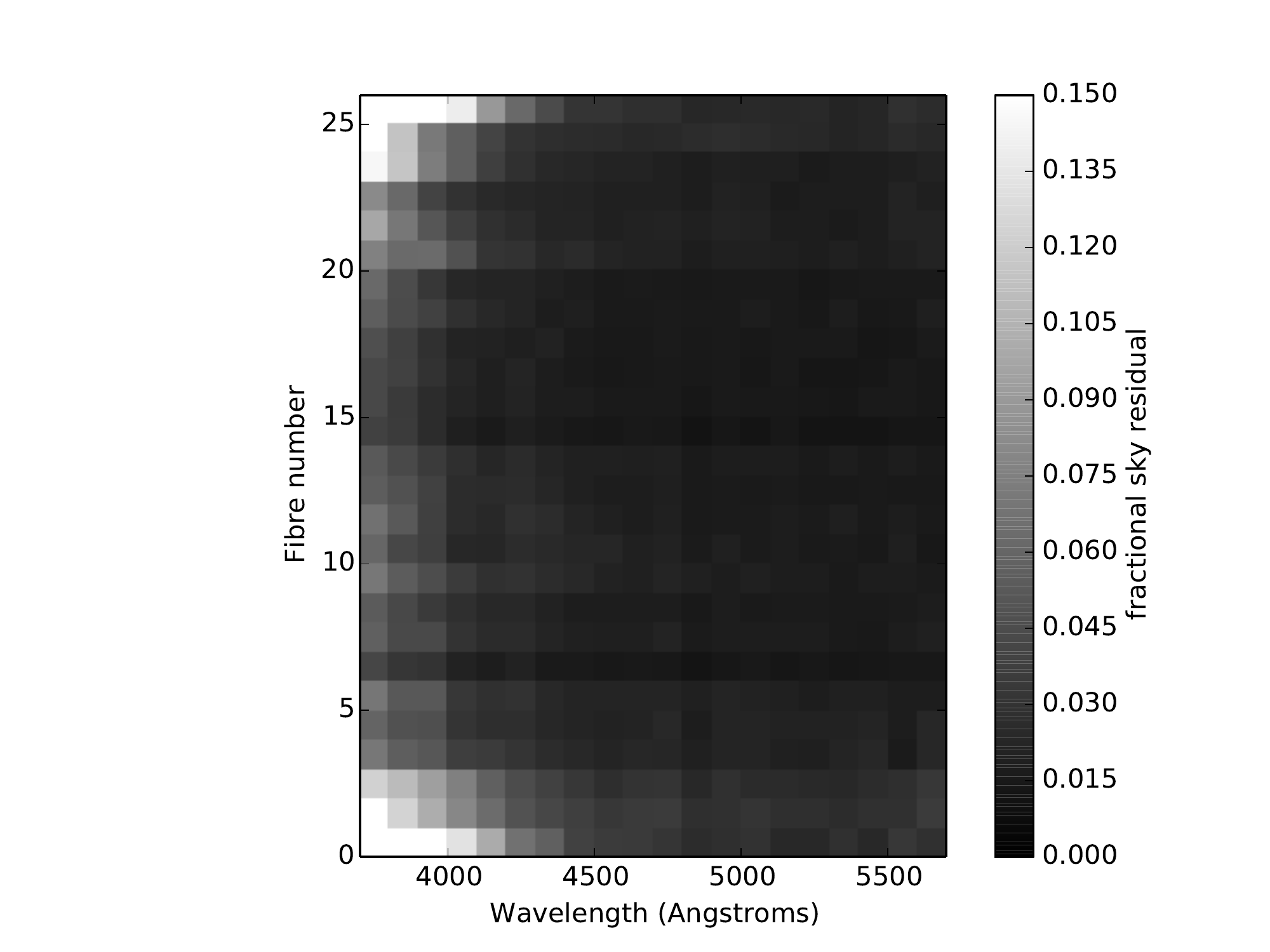}%
  \includegraphics[width=0.5\linewidth]{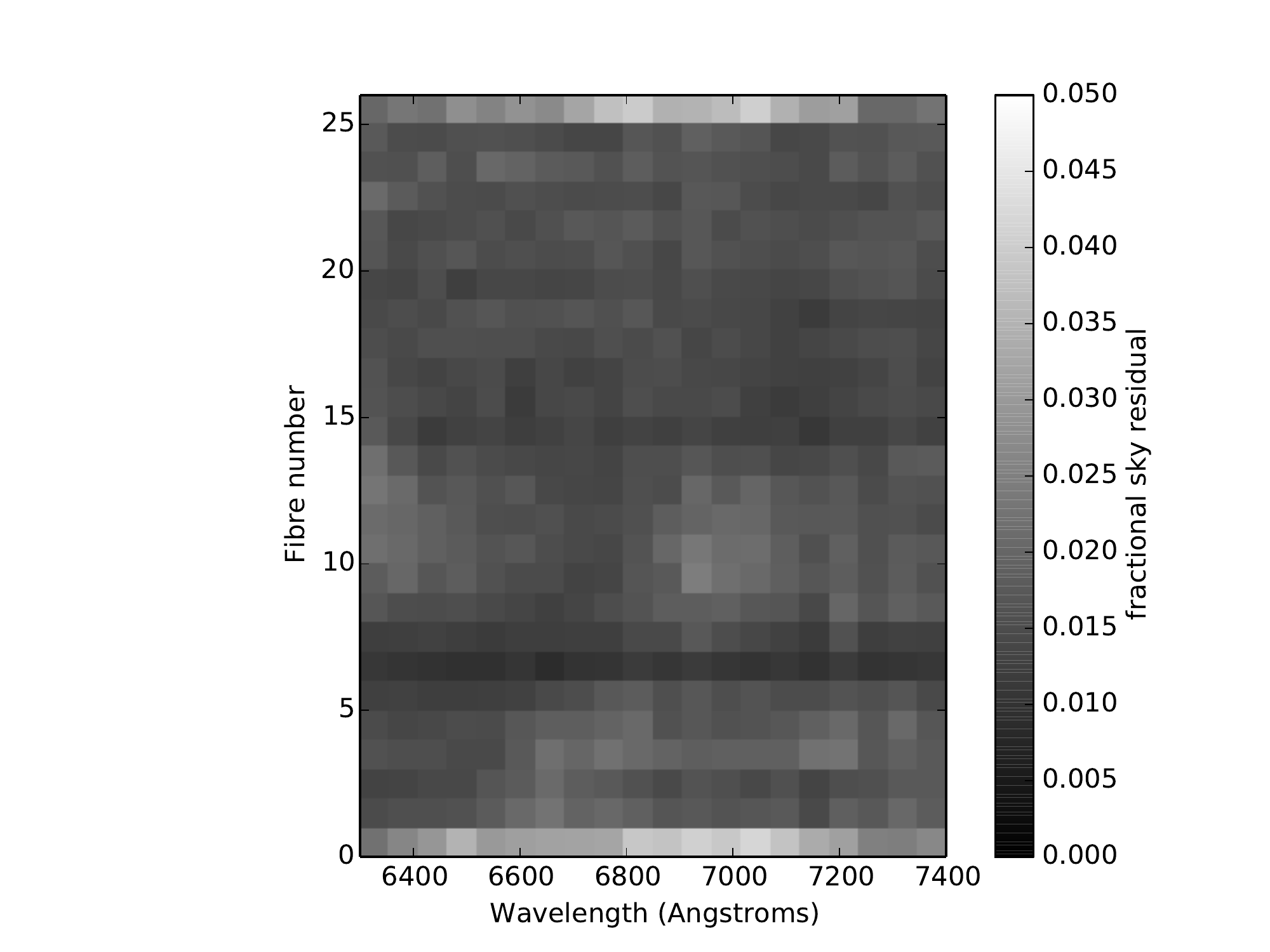}
  \caption{ %
    \label{fig:skysubmap}%
    The median fractional sky subtraction residuals as a function of
    wavelength and fibre number for SAMI sky fibres in the blue (left)
    and red (right) arms of the spectrograph.  The sky fibres are
    regularly spaced along the SAMI slit, so that sky fibre number
    also corresponds to approximate location on the AAOmega CCDs. If
    the sky subtraction was perfect, these residuals would be
    zero---instead they indicate the likely sky-subtraction residuals
    in science fibres adjacent to these sky fibres.  For each sky
    fibre shown, the spectral direction is sub-divided into 20 uniform
    bins, and the residual flux is summed in each of these bins,
    before determining the median residual (across different observed
    frames).  This reduces the impact of shot noise on the residual
    estimate and allows us to see systematic variations in sky
    subtraction.  A strong increase in the residual in the left hand
    corners of the blue CCD are particularly apparent.  Note the
    difference in grey-scale between the two images. %
  }
\end{figure*}

Residuals after subtracting sky-continuum may instead arise from
scattered light in the spectrograph. The residuals are shown as a
function of wavelength and sky fibre number in Figure~\ref{fig:skysubmap}.
To clarify the impact of sky subtraction errors, we sum the residual
flux in wavelength bins (20 uniform bins per spectrograph arm).  The
sum reveals sky residuals that would otherwise be dominated by CCD
read noise and photon counting errors in a single 0.5 to 1\AA-wide
wavelength channel.  Figure~\ref{fig:skysubmap} shows that across most
of both the blue and red arm CCDs, residuals of the sky-continuum
subtraction are $\sim1$ percent.  However, a strong residual appears
at the short wavelength corners of the blue CCD.  This is due to a
ghost in the spectrograph caused by a double bounce between the CCD
and air-glass surfaces of the AAOmega camera corrector lens (Ross
Zhelem, private communication).  The ghost results in poor fitting of
the fibre profiles, which in turn results in poor extraction and then
sky subtraction.  A solution to this using twilight sky flats to
generate fibre profiles has now been developed, but has not been
applied to the data in DR1.
%                              (155 words)

\subsubsection{Point spread function}
\label{sec:psf}

\begin{figure}
  \centering
  \includegraphics[width=0.8\linewidth]{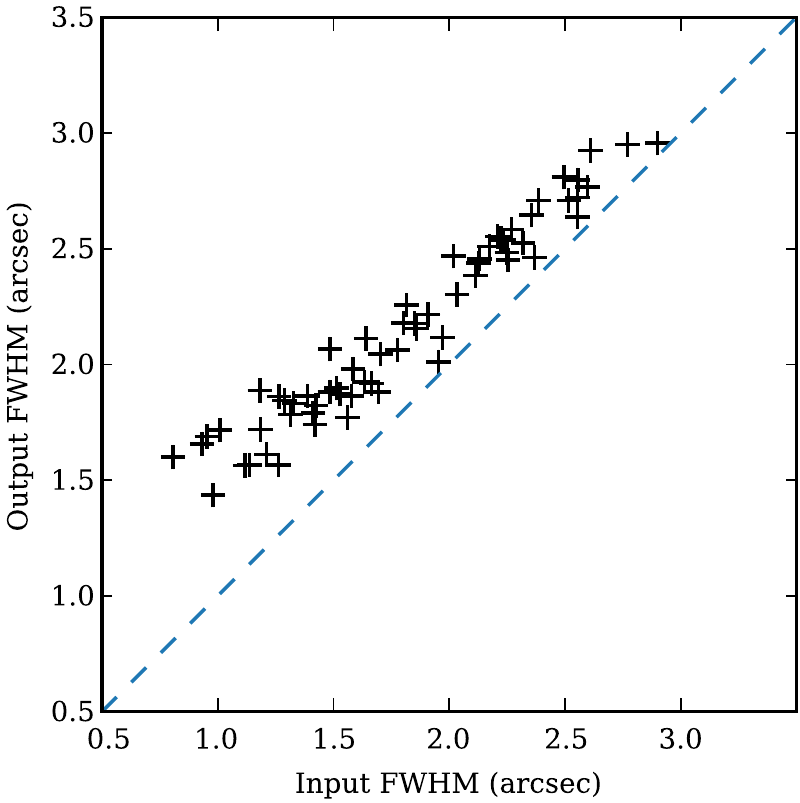}
  \caption{%
    \label{fig:fwhm-change} %    
    Comparison of the FWHM of the reconstructed secondary standard
    images (Output FWHM) versus the mean FWHM of the individual
    dithered exposures (Input FWHM). FWHMs are from Moffat-profile
    fits. The dashed blue line is the 1:1 relation. The output FWHM is
    typically larger than the input by $10\%$. Also shown is the
    histogram of the Output FWHM. The mean FWHM for DR1 is
    $\n{meanSeeing}$~arcsec, and the standard deviation is
    \n{meanSeeingStd}~arcsec. %
  }
\end{figure}

The spatial PSF is measured by fitting a Moffat function to the
reconstructed image of the secondary-standard star in each SAMI field.
SAMI fibres have diameter $1.6 \; \mathrm{arcsec}$, therefore in
seeing $\lesssim 3 \; \mathrm{arcsec}$, the PSF in the individual
dithered exposures is under-sampled. Stacking images introduces
additional uncertainty from mis-alignment of the seven frames
\citep[figure~15 of][]{2015MNRAS.446.1567A}, and from combining
exposures with slightly different seeing.  Therefore, the PSF of the
final spectral cube is degraded from the PSF of the individual
frames. In Figure~\ref{fig:fwhm-change} we compare the FWHM of the
reconstructed stellar image (output FWHM) to the mean FWHM of the
individual exposures (input FWHM). For small input FWHM ($\approx 1 \;
\mathrm{arcsec}$), output FWHM increases by $50\%$.  This regime is
likely dominated by PSF under-sampling.  When input FWHM exceeds
$\approx 1.5 \; \mathrm{arcsec}$, output FWHM is typically $10\%$
larger.  No stars have FWHM $> 3.0 \; \mathrm{arcsec}$ as such data is
excluded by a quality control limit. In summary, DR1 spectral cubes
have a mean PSF of $\n{meanSeeing}$~arcsec (FWHM).

\subsubsection{Flux Calibration}
\label{sec:dq-flux-cal}

The relative flux calibration as a function of wavelength in DR1 is
consistent with that in the EDR.  By comparing SAMI data with SDSS
$g-$ and $r-$band images, \citet{2015MNRAS.446.1567A} showed that SAMI
derived $g-r$ colours have 4.3 percent scatter, with a systematic
offset of 4.1 percent, relative to established photometry.

\begin{figure}
  \centering  
  \includegraphics[width=0.8\linewidth]{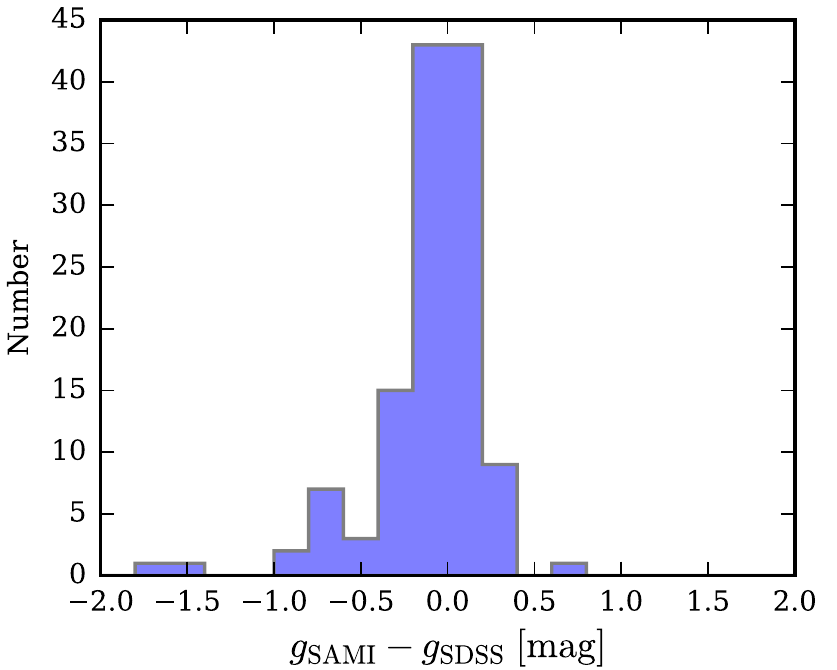}
  \caption{ %
    \label{fig:gal-mag} %
    Distribution of the offset in g-band magnitude between the SAMI
    spectral cube and SDSS magnitudes. Only galaxies with
    $\texttt{petroR50\_r} < 2\,\mathrm{arcsec}$ are included. The mean
    offset is $-0.07 \, \mathrm{mag}$ and the standard deviation is
    $0.22 \, \mathrm{mag}$. %
  }
\end{figure}

To test the absolute flux calibration, Figure~\ref{fig:gal-mag} shows
the distribution of \textit{g}-band magnitude differences between the
SAMI galaxies and the corresponding Petrosian magnitudes from SDSS. To
avoid aperture losses and extrapolations, the distribution is only
shown for the 127 SAMI galaxies having Petrosian half-light radius
($\texttt{petroR50\_r}$) $ < 2 \, \mathrm{arcsec}$ from SDSS. The
median offset is $-0.07 \, \mathrm{mag}$, and the standard deviation
is $0.22 \, \mathrm{mag}$.  This is an improvement over the standard
deviation of $0.27 \, \mathrm{mag}$ in the EDR.  As pointed out by
\citet{2015MNRAS.446.1567A}, there is a $0.14 \, \mathrm{mag}$ scatter
between SDSS Petrosian and model magnitudes for our sample, so a
considerable fraction of the $0.22 \, \mathrm{mag}$ scatter is likely
due to the inherent limitations in galaxy photometry.

\subsubsection{WCS and Centring of Fibre Bundles in Cubes}
\label{sec:wcs-accuracy}

The accuracy of the WCS is limited by the stability and accuracy of
the single Gaussian fit on the observation chosen as the reference
(typically the first frame, see Section~\ref{sec:spectral-cubes} and
section~5.2 of \citealt{2015MNRAS.446.1551S}).  By fitting to the
individual observed galaxies we lose some robustness. However, we
minimize the impact of mechanical errors (plate manufacturing,
movement of the connectors within the drilled holes, and uncertainty
of the bundle positions) on the WCS accuracy.  Examining the data, we
have identified three possible failure modes of our approach:
\begin{itemize}
\item The fit may identify a bright star within the field of view of
  the hexabundle instead of the galaxy of interest. Examples include
  galaxies 8570 and 91961. 
\item The catalogue coordinate may not correspond to a peak in the
  surface brightness of the object, such as one with a very disturbed
  morphology, or for objects where the catalogue coordinate has been
  intentionally set to be between two galaxies (galaxies with
  BAD\_CLASS=5 in the target catalogue), see \cite{2015MNRAS.447.2857B}
  for details. Examples include galaxy 91999.
\item Finally, the circular Gaussian distribution may not represent
  the true flux distribution well, leading to some instability or bias
  in the fit result. Examples include large, extended galaxies such as
  514260.
\end{itemize}
In these cases the WCS origin may not be very
accurate, and the hexabundle field of view may not be well centred in the
output spectral cube. 

We carry out two tests to characterise uncertainties in the WCS.  The
first is an internal check that considers offsets at different stages
of the alignment process to constrain the expected WCS
uncertainties. The second cross-correlates the reconstructed SAMI
images with SDSS broad-band images to measure the offset between SAMI
and SDSS coordinates. These two tests, which we detail in
the following paragraphs, suggest that the WCS accuracy is $\lesssim
0.3$~arcsec for most galaxies, except for the failures noted
above.

\begin{figure}
  \centering  
  \includegraphics[width=1\linewidth]{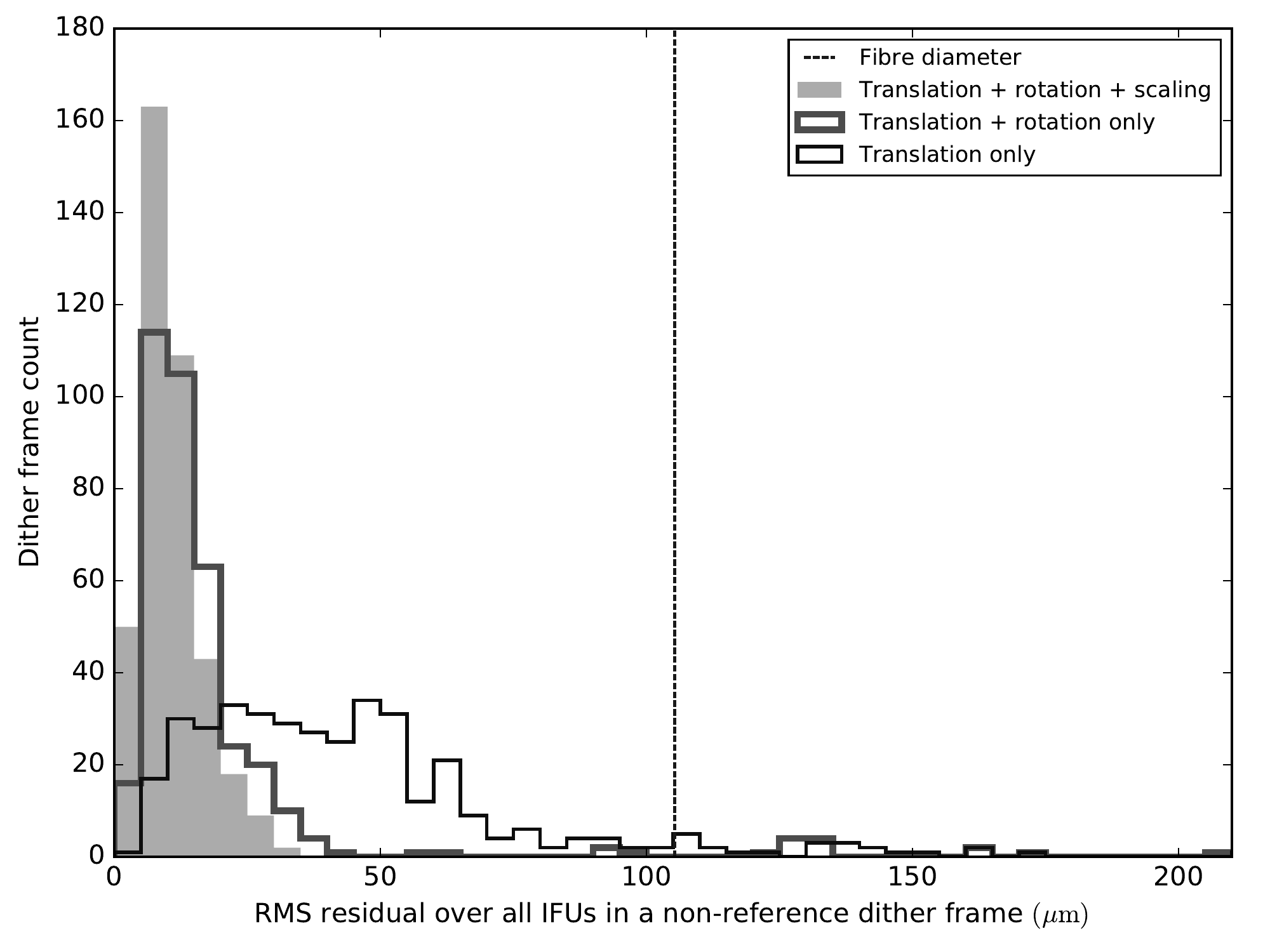}
  \caption[WCS Centroid Offsets]{%
    \label{fig:wcs-centroid2}%
    Histograms of the residuals after aligning dither frames.  The
    alignment attempts to simultaneously place bundle centroids of all
    \acp{IFU} in a dither onto the centroids of the `reference' dither
    frame.  Distributions of residuals are shown for transformations
    with only a translation, a translation and a rotation, and the
    full transformation of a translation, a rotation, and a scaling.%
  }
\end{figure}

The internal tests to examine WCS uncertainties use alignment offsets
to infer bounds on the typical size of the WCS uncertainties.  The
first dither pointing of an observation aims to centre each galaxy in
its bundle.  The dither-alignment transformation aligns the galaxy
centroid positions in a dither with the galaxy centroid positions in
the first (`reference') frame of an observation.
Figure~\ref{fig:wcs-centroid2} shows the RMS of the residuals for all
bundles in a dither after the dither was aligned with the reference
frame.  The residuals are shown for transformations that are
translation-only, translation and rotation, and using the full
transformation of a translation, rotation and scaling.  At least
translation is necessary because the dithers are deliberately
spatially offset.  However rotation is also important in aligning the
dither frames to the centre of the cubes as the SAMI instrument plate
holder has a small ($\sim0.01$ degrees) bulk rotation away from its
nominal orientation.  This rotation suffices to generate offsets from
the nominal bundle centres of up to $\sim1$ arcsec at the edge of the
field of view.  A further improvement is gained using the modification
of the plate scale, due to differential atmospheric refraction causing
small positional shifts over the course of an observation.  The mean
RMS of ${\sim}11$~$\micron$ (0.16~arcsec) for the full transformation
reflects how accurately the data are spatially combined for a typical
galaxy and hence provides a lower limit to the WCS uncertainty.

\begin{figure}
  \centering  
  \includegraphics[width=0.7\linewidth,trim=30 0 10 0]{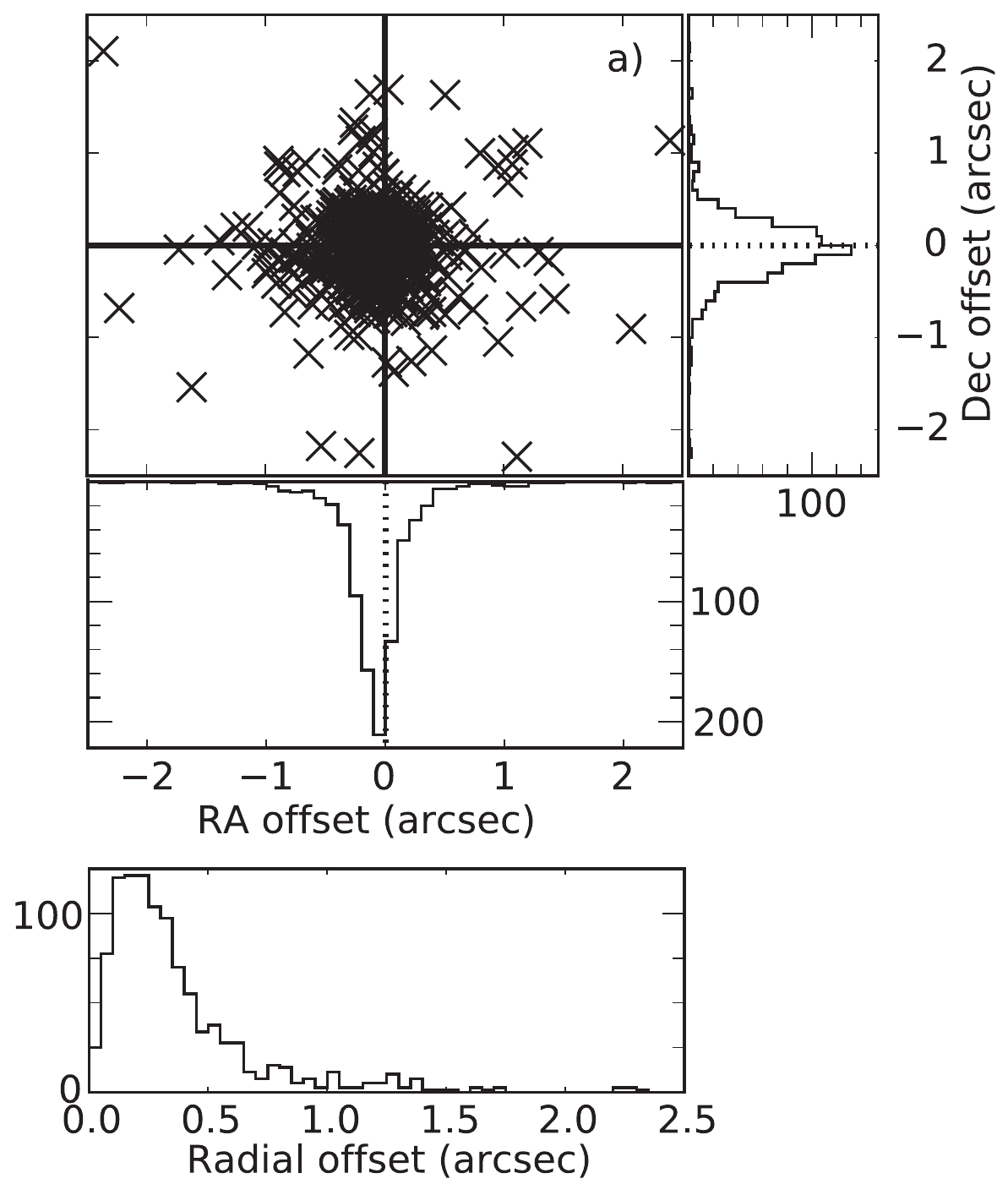}
  \caption[WCS test]{ %
    \label{fig:wcs_test} %
    The difference between SAMI and SDSS astrometric solutions based
    on cross-correlation of images.  (top) The distribution of RA and
    declination differences between SAMI and SDSS, with histograms of
    the differences in declination and right ascension along the
    axes. (bottom) Histogram of the distribution of differences in
    radial offset. %
  }
\end{figure}

The cross-correlation test of the WCS accuracy compares the spatial
flux distribution of the final, reconstructed SAMI cubes to SDSS
$g$-band images.  Each cube is multiplied by the SDSS-$g$-band-filter
response and then summed spectrally.  The resulting image is then
cross-correlated with an SDSS $g$-band image. These SDSS
images are centred on the expected coordinates of the galaxy (based on
the GAMA input catalogue), are $36\times36$\,arcsec in size, and have
been re-sampled to the same 0.5~arcsec pixel scale as the SAMI cubes.
The cross-correlation offset (measured using a fit to the peak in the
cross-correlation image) is then the difference between the SAMI WCS
and the SDSS WCS. These differences are shown in
Figure~\ref{fig:wcs_test}.  Outliers in most cases are caused
by the cross-correlation centring on bright stars that are present in
the SDSS image, but not in the SAMI field of view. Visual checks of
outliers also identified five galaxies with gross errors in their SAMI
cube WCS, caused by the data reduction centroiding on a bright star in
the SAMI field of view rather than the target galaxy (catalogue IDs
8570, 91961, 218717, 228104 and 609396). When outliers are removed
using an iterative 5$\sigma$ clipping (that removes 7.7 per cent of
coordinates), the mean of the remaining differences is
$-0.074\pm0.020$\,arcsec in right ascension and
$-0.048\pm0.037$\,arcsec in declination.  The root-mean-square scatter
is 0.18\,arcsec in right ascension and 0.27\,arcsec in
declination. This test suggests a typical radial WCS error of
0.32\,arcsec. 

Given that the result of the measurement of the WCS uncertainty in the
cross-correlation test is consistent with the bounds suggested by the
internal tests, we expect that it is representative of the actual
uncertainty in our WCS for most targets. The targets subject to one of
the failures mentioned above will have a much larger error in their
WCS (no attempt has been made to correct these failures).

\subsection{Impact of aliasing from sampling and DAR on SAMI data}

The combined effects of DAR and limited, incomplete spatial sampling
can cause the PSF of IFS data to vary both spatially and spectrally
within a spectral cube, an effect we call ``aliasing''.  We describe
this in the Appendix, but \citet{2015AJ....150...19L} also provide an
excellent discussion. Aliasing can cause issues in comparing widely
separated parts of the spectrum on spatial scales comparable to, or
smaller than, the size of the PSF. Examples are spectral colour and
ratios of widely spaced emission lines. We therefore check the impact
of aliasing on our data and discuss options for reducing this impact.

\begin{figure}
  \centering  
  \includegraphics[width=1.0\linewidth]{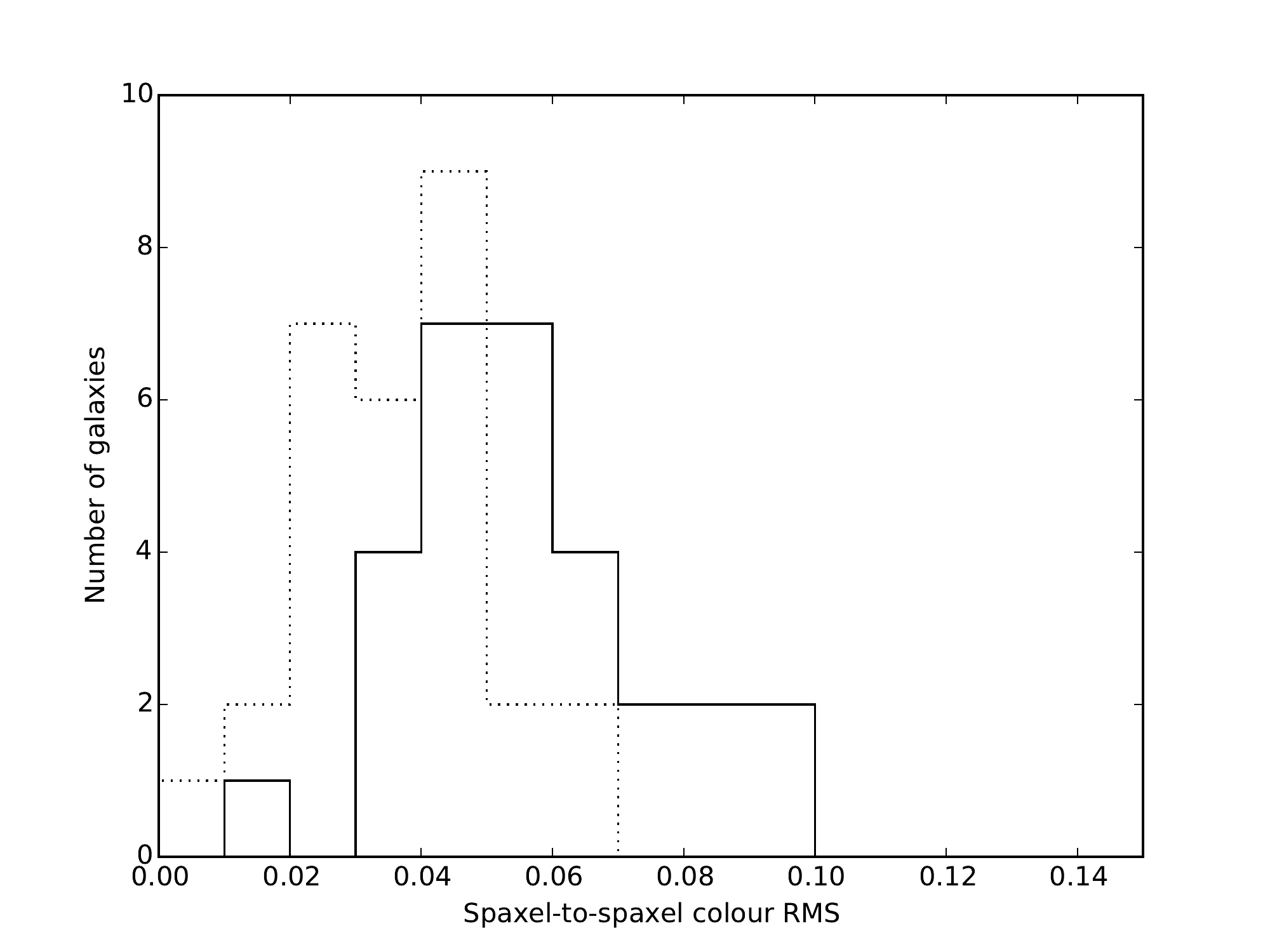}
  \caption[Spaxel RMS]{%
    Histogram of RMS scatter in colour between spaxels that are
    $0.5\times0.5$ (solid line) or $1.0\times1.0$ (dotted line) arcsec in size.
    Galaxies tested are chosen to be passive with uniform colour and
    the RMS is calculated independently for each galaxy. %
  }
  \label{fig:dar_col_rms}
\end{figure}

To test the impact of aliasing in SAMI data, we check the variation in
colour within galaxies expected to have uniform colour across their
extent. Uniform colour galaxies are chosen to be passive (no
significant emission lines) and to have weak (or flat) stellar
population gradients.  Using only spaxels in the blue SAMI cubes that
have a median S/N $>15$, we smooth them with a Gaussian kernel in the
spectral direction ($\sigma=15$\AA) to reduce noise, and then sum the
flux in two bands at wavelengths 3800--4000\AA\ and
5400--5600\AA. These bands are chosen to be narrower than typical
broad-band filters, but be more sensitive to the size of the aliasing
effects (see Appendix).  For each galaxy we then estimate the RMS
scatter in the colour formed by the ratio of the flux in these two
bands.  Figure~\ref{fig:dar_col_rms} shows the distribution of RMS
scatter measurements in the spaxel-to-spaxel spectral colour for 29
galaxies.  For the default $0.5\times0.5$-arcsec spaxels (solid line
in Figure~\ref{fig:dar_col_rms}) the median scatter is 0.052 and the
5th--95th percentile range is $0.033-0.093$.  Summing spaxels
$2\times2$ within the cubes so that we have $1.0\times1.0$-arcsec
spaxels (dotted line in Figure~\ref{fig:dar_col_rms}) leads to a
reduced RMS with median value of 0.035 and the 5th--95th percentile
range is $0.012-0.061$.  The reduction in scatter when the data are
binned to larger spaxels is consistent with the scatter being caused
by aliasing in DAR re-sampling.

Aliasing from DAR re-sampling can also affect line-ratios. The ratio
of the \Ha\ and \Hb\ emission lines is typically used to estimate dust
attenuation. Variations in the PSF at these two wavelengths
causes the ratio to reflect not only the true ratio of the two lines,
but also the difference in the PSF between the two wavelengths. The
later effect will be most pronounced where there is a sharp change in
flux with spatial position in either of the two lines (such as near an
unresolved \ion{H}{2} region). In such a region, there will be
variations pixel-to-pixel (smaller than the PSF) that are larger than
would be indicated by the variance information of the data
alone.

One possible method for reducing the impact of aliasing on SAMI data
is to smooth it.  For example, smoothing the \Ha-\Hb\ line ratio map
by a 2D Gaussian kernel of Gaussian-$\sigma$ of 0.5~arcsec (one
spatial pixel) and truncated to $5\times5$ pixels removes most of the
variation caused by aliasing without greatly affecting the output
spatial resolution.  This smoothing brings the noise properties of the
\Ha-\Hb\ line ratio into agreement with Gaussian statistics and
significantly reduces variation in the normalised spectra for
(point-source) stars.  The best choice for the smoothing kernel
$\sigma$ probably ranges between 0.2 and 1~arcsec, depending on the
science goal and the level of DAR aliasing associated with the galaxy
properties and observational conditions.  Smoothing should only be
necessary when no other averaging is implicit in the analysis
(e.g. smoothing is not necessary for measuring radial gradients).

Alternative data reconstruction schemes may reduce the effects of
aliasing from the DAR re-sampling.  Smoothing options are discussed
further in A. Medling et~al.\ (submitted) as they pertain to the
emission-line Value Added Products (described briefly in
Section~\ref{sec:vap}). In general only results that depend on the
highest possible spatial resolution are likely to be sensitive to
aliasing.

\section{Emission-Line Physics Value-Added Data Products}
\label{sec:vap}

\begin{figure*}[p]
  \centering  
  \includegraphics[width=0.9\linewidth]{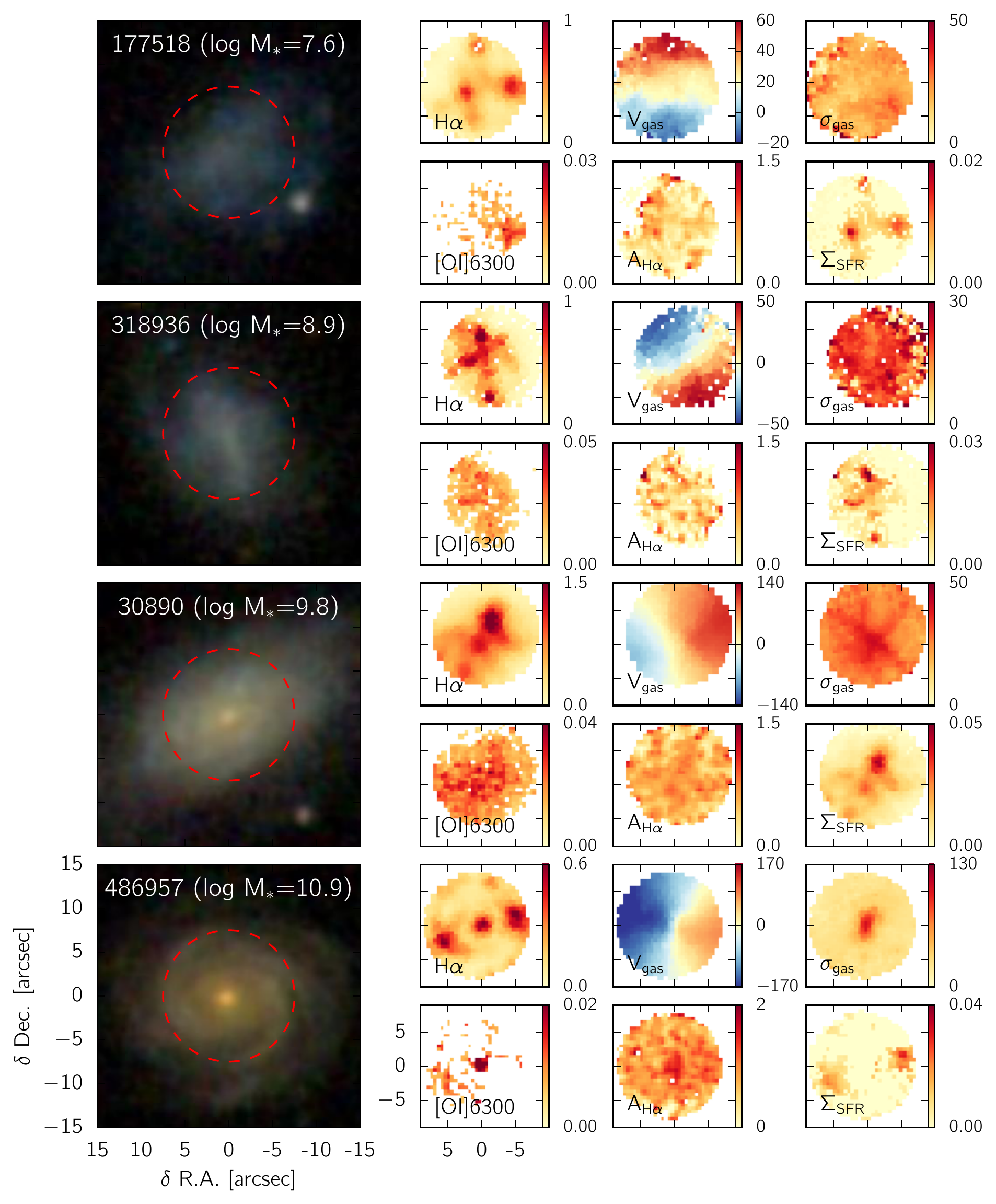}
  \caption[Overview of star-formation value-added products]{%
    \label{fig:vap-overview}%
    Examples of value-added products for four galaxies spanning the
    range of stellar masses included in our DR1. The red dashed
    circles on the SDSS 3-colour images (left) indicate the radius of the SAMI
    fibre bundle. The small panels show the various
    value-added products: H$\alpha$ flux, gas velocity ($v_{gas}$), gas
    velocity dispersion ($\sigma_{gas}$), [\ion{O}{1}]$\lambda 6300$ flux,
    H$\alpha$ attenuation correction factor ($\rm A_{H\alpha}$), and
    star formation rate surface density $\Sigma_{\rm SFR}$ maps. The
    units are $\rm 10^{-16}~ergs~s^{-1}~cm^{-2}~arcsec^{-2}$ for the
    flux maps, $\rm km~s^{-1}$ for the kinematic maps, magnitude for
    $\rm A_{H\alpha}$, and $\rm M_\odot~yr^{-1}~kpc^{-2}$ for
    $\Sigma_{\rm SFR}$. %
  }
\end{figure*}

With the Core Data Products described above, our DR1 also
includes Value-Added Products based on the ionized-gas emission lines
in our galaxies.  We provide fits for eight emission lines from five
ionisation species, maps of Balmer extinction, star-formation masks,
and maps of star-formation rate for each galaxy. Examples of these
products are shown in Figure~\ref{fig:vap-overview} for a selection of
galaxies spanning the range of stellar masses in DR1.

\subsection{Single- and multi-component emission-line fits}
\label{sec:emission-line-fits}

We have fit the strong emission lines (\oii\ 3726,3729, \Hb, \oiii\
4959,5007, \oi\ 6300, \nii\ 6548,6583, \Ha, and \sii\ 6716,6731) in
the spectral cubes with between one and three Gaussian profiles. We
fit with the {\tt LZIFU} software package detailed in
\cite{2016ApSS.361..280H}.  These fits include corrections for
underlying stellar-continuum absorption.  \texttt{LZIFU} produces both
a single component fit and a multi-component fit for each spatial
pixel of the spectral cube. The latter fits select the optimum number
of kinematic components in each spatial pixel.

All lines are fit simultaneously across both arms of the
spectrograph. The blue and red spectral cubes have FWHM spectral
resolutions of \n{specfwhmblue} and \n{specfwhmred},
respectively. Assuming that the kinematic profiles are consistent for all
lines, the higher resolution in the red helps to constrain the fits in
the blue, where individual kinematic components may not be resolved.

{\tt LZIFU} first fits underlying stellar continuum absorption using the
penalized pixel-fitting routine \citep[PPXF;][]{2004PASP..116..138C},
then uses {\tt MPFIT} \citep[the Levenberg-Marquardt least-square
method for IDL;][]{2009ASPC..411..251M} to find the best-fit
Gaussian model solution.

Our continuum fits combine template spectra of simple stellar
populations from the Medium resolution INT Library of Empirical
Spectra \citep[MILES,][]{2010MNRAS.404.1639V}. These spectra are based
on the Padova isochrones \citep{2000A&AS..141..371G}. The selected
templates have four metallicities ($[M/\textrm{H}] = -0.71,\ -0.40,\
0.0,\ +0.22$) and 13 ages (logarithmically spaced between 63.1 Myr and
15.8 Gyr). In fitting the template spectra to our observed data,
Legendre polynomials (orders 2-10) are added (not multiplied) to
account for scattered light and other possible non-stellar emission
within the observed spectral cubes, and a reddening curve parametrised
by \cite{2000ApJ...533..682C} is applied. Note that the MILES
templates have slightly lower spectral resolution than the red arm of
our spectra; therefore, in low-stellar-velocity-dispersion galaxies
($\sigma<30\kms$), the template may under-estimate the \Ha\
absorption.  To account for this and other systematic errors from
mis-matched templates, we calculate the expected uncertainty in the
Balmer absorption from the uncertainty in stellar-population age as
measured from the size of the D$_{\rm n}$4000 break.  This uncertainty
is added into the Balmer-emission-flux uncertainty in quadrature.
%                              (163 words)

\begin{figure*}
  \centering  
  \includegraphics[width=0.9\linewidth]{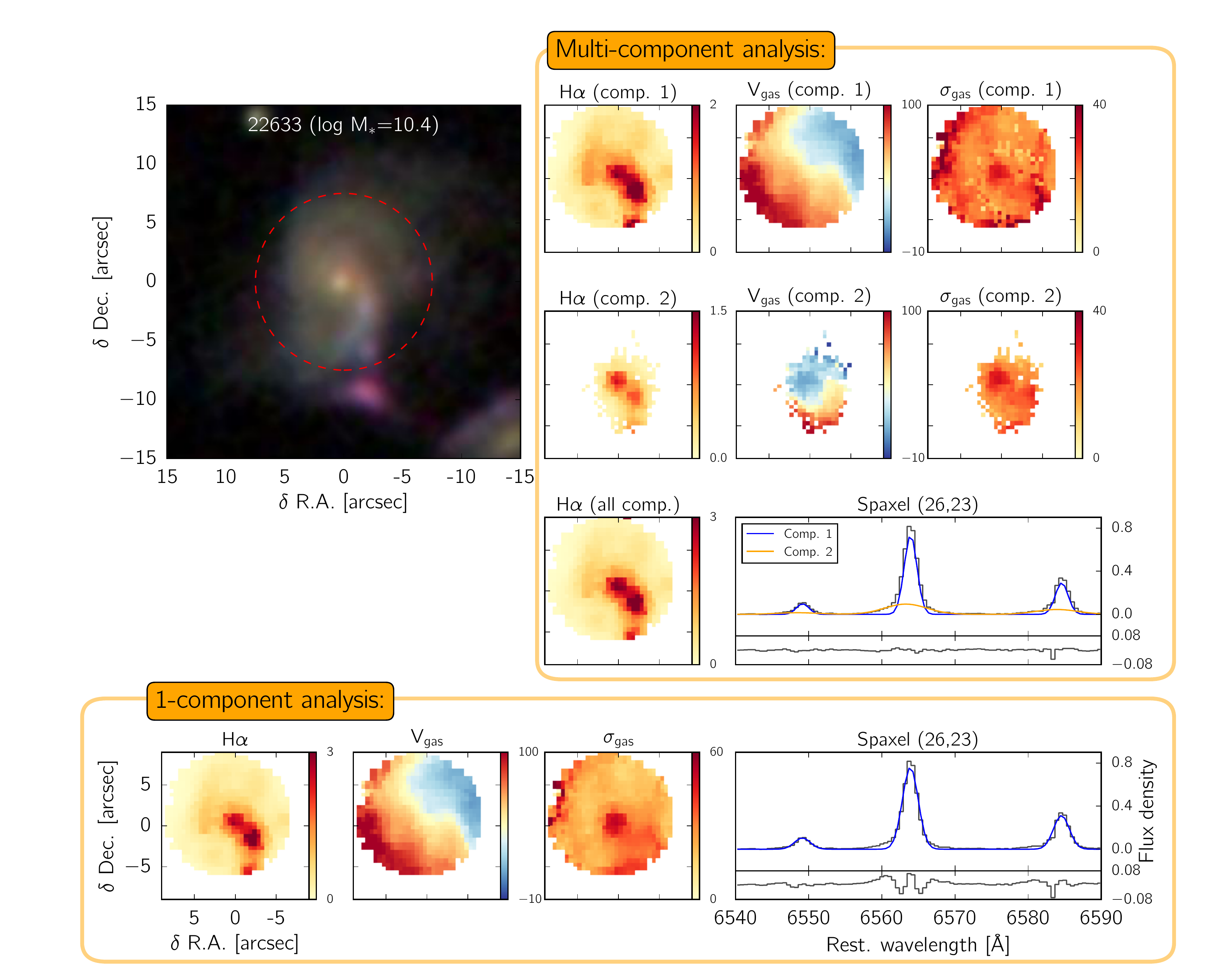}
  
  \caption[Example of Multi-component emission line fits]{%
    \label{fig:multicom}%
    An example comparing multi-component analysis to single-component
    analysis. The red dashed circle in the SDSS colour image (top
    left) indicates the SAMI fibre bundle. With the multi-component
    analysis, we demonstrate that there are two distinct kinematic
    components in GAMA~22633. The two kinematic components show
    different velocity dispersions in the spectral fitting panel. The
    two components also show different H$\alpha$ distribution
    (H$\alpha$ maps for comp. 1 and 2) and velocity structures
    ($v_{gas}$ maps for comp.1 and 2.). The nature of the second
    kinematic component cannot be determined with the 1-component
    analysis that only captures the more dominant narrow kinematic
    component (bottom row). Including the second kinematic component
    is necessary to properly model the line profile and reduce the
    residual.  %
  }
\end{figure*}

Each emission line in each spaxel is fit separately with one, two, and
three Gaussian components. In each case, a consistent velocity and
velocity dispersion are required for a given component across all
lines.  For each galaxy, DR1 includes two sets of fits: one that uses
a single Gaussian for each line in each spatial pixel (``single
component''), another that includes one to three components for each
spatial pixel (``recommended components''). Examples of these two fits
are shown in Figure~\ref{fig:multicom}. For the fits with recommended
components, the number of fits included for each spatial pixel is
chosen by an artificial neural network trained by SAMI Team members
\citep[\texttt{LZComp},][]{2016arXiv160608133H}. For the recommended
components, we also require that each component has S/N $\ge5$ in \Ha;
if this condition is not met, we reduce the number of components until
it does.
%                              (143 words)

The single-component fits include eight maps of line fluxes, and a map
each of ionized gas velocity and velocity dispersion. The \oii\
3726,3729 doublet is summed because the blue spectral resolution
prevents robust independent measurements of it's components. Flux maps
of \oiii\ 4959 and \nii\ 6548 are omitted because they are constrained
to be exactly one-third of \oiii\ 5007 and \nii\ 6583, respectively.
%                              (78 words)

The recommended-component fits include maps of the total line fluxes
(i.e.\ the sum of individual components) for each emission line.
Additionally, for the \Ha\ line, three maps show fluxes of the
individual fit components, and there are three maps each of the
velocity and velocity dispersions, which correspond to the
individual components of the H$\alpha$ emission line.  The maps
showing individual components of \Ha\ flux, velocity, and velocity
dispersion are ordered by component width, i.e.  first corresponds to
the narrowest line and third to the widest. Where there are fewer than
three components, higher numbered components are set to the floating
point flag \texttt{NaN}, as are all maps without a valid fit.

Figure~\ref{fig:poster-plot} illustrates the value of the emission
line fits and the richness of our DR1. It shows how the nature of gas
emission changes within galaxies as a function of their stellar mass
and star-formation rate. At lower stellar masses, emission is driven
by star formation, and the gas typically has lower metallicity, which
is represented by lower \nii/\Ha\ ratios (blue). At higher stellar
masses, low-star-formation-rate galaxies often host AGN, often
resulting in the prominent peak in \nii/\Ha\ ratio at the centre of
the galaxy (red).

Our DR1 includes total-flux model spectral cubes
(continuum model plus all fitted emission lines) for direct comparison
with the spectral cubes, and maps of quality flags to highlight
issues such as bad continuum fits or poor sky subtraction.

\subsubsection{Accuracy of GAMA redshifts and systemic velocities from emission line fits}

LZIFU derived velocities are with reference to the catalogued GAMA
redshifts that are listed in the SAMI input catalogue
\citep[see][]{2015MNRAS.447.2857B}.  The GAMA redshifts are on a
heliocentric frame and sourced from various surveys such as the main
GAMA spectroscopic program \citep{2013MNRAS.430.2047H}, SDSS
\citep{2000AJ....120.1579Y}, and 2dFGRS \citep{2001MNRAS.328.1039C}.
To check the velocity scale of the SAMI cubes, we construct aperture
spectra by summing across an $1R_{\rm e}$ ellipse.  For SAMI cubes
that do not extend to $1R_{\rm e}$, we sum over the whole SAMI cube.
Each aperture spectrum is then fit with LZIFU using exactly the same
process as the individual cube spaxels.

\begin{figure}
  \centering  
  \includegraphics[width=\linewidth]{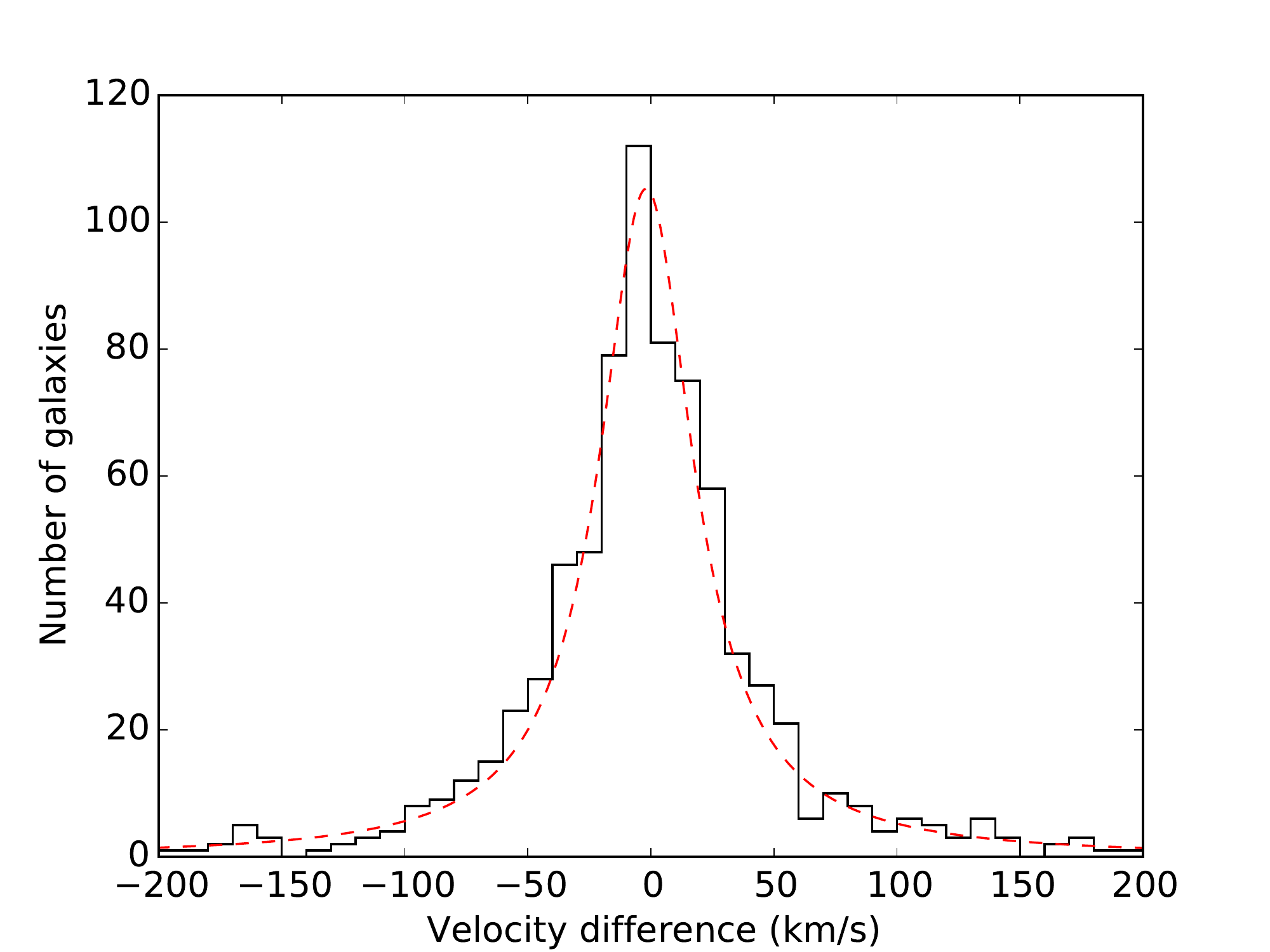} 
  \caption{%
    \label{fig:vel_diff}%
    The distribution of rest-frame velocity differences between
    redshifted catalogued by GAMA and those from LZIFU for SAMI
    $1R_{\rm e}$ aperture spectra, both corrected to the
    heliocentric reference frame.  The red dashed line is a Lorentzian
    fit to the distribution. %
  }
\end{figure}

Figure~\ref{fig:vel_diff} shows the velocity difference between the
assumed GAMA redshifts and that measured in the aperture spectra.  The
median difference is $-1.6$\kms\ and a robust $1\sigma$ range based on
the 68--percentile range is $43.9$\kms.  The GAMA redshifts used in
the SAMI input catalogue were measured using the \textsc{runz} code,
and GAMA reports an error on individual \textsc{runz}-derived
emission-line redshifts of $33$\kms\ from repeat observations
\citep[using a robust 68--percentile range;][]{2015MNRAS.452.2087L}.
By subtracting the two in quadrature, we estimate an intrinsic scatter
of 35\kms\ for our DR1.  This number is an upper limit to the true
scatter in SAMI velocity measurements, because it also accounts for
differences due to the spatial distribution of \Ha.  For example, with
single-fibre observations targeting a location that is not the
dynamical centre of a galaxy, or the SAMI aperture spectrum being
dominated by strong \Ha\ flux in the outer parts of galaxies in some
cases.

The distribution of velocity differences is well described by a
Lorentzian distribution, as found by \cite{2015MNRAS.452.2087L} for
the GAMA velocity uncertainties.  The best fit Lorentzian is shown by
the red dashed line in Figure~\ref{fig:vel_diff}.  The galaxies in the
wings of the distribution of velocity differences tend to be those
that have lower S/N ratio in the emission line flux.

\subsection{Star Formation Value-Added Products}
\label{sec:sf-vap}

Included with DR1 are value added products necessary for understanding
the spatially-resolved star formation. These are:
\begin{itemize}
\item Maps of \Ha\ extinction: these are derived by assuming a Balmer
  decrement (\Ha/\Hb~ratio), unphysical ratios have extinction
  corrections set to 1 (no correction). Uncertainties in the
  extinction correction are also provided.
\item Masks classifing each spaxel's total emission-line flux as
  `star-forming' or `other': these are derived using the line-ratio
  classification scheme of \cite{2006MNRAS.372..961K}.
\item Maps of star-formation rate: these are derived from \Ha\
  luminosities and include the extinction and masking above. The
  conversion factor used is $7.9\times 10^{-42} \Msunyr
  (\ergsec)^{-1}$ from \citet{1998ApJ...498..541K}, which assumes a
  Salpeter initial mass function \citep{1955ApJ...121..161S}.
\end{itemize}
These data products will be described in detail in a companion paper by
Anne Medling, et~al.

\section{Online Database}
\label{sec:database}

The data of this Release are presented via an online database
interface available from the Australian Astronomical Observatory's
Data Central\footnote{Data Central's URL is
  \href{http://datacentral.aao.gov.au}{http://datacentral.aao.gov.au}}. Data
Central is a new service of the Observatory that will ultimately
deliver various astronomical datasets of significance to Australian
research. Users of the service can find summary tables of the galaxies
included in our DR1, browse the data available for individual
galaxies, and visualize data interactively online. The service
provides for downloading individual and bulk data sets, and a
programmatic interface allowing direct access to the data through the
HTTP protocol. Also provided are extensive documentation of DR1, the
individual datasets within it, and the formatting and structure of the
returned data.

Data Central presents data in an object-oriented, hierarchical
structure. The primary entities of the database are astronomical
objects, such as stars or galaxies. These entities have various
measurements and analysis products associated with them as
properties. For example, each galaxy in our DR1
is an entity in the database, with properties such as red and blue
spectral cubes, LZIFU data products, and star-formation maps. In
future, these galaxies may also have data from other surveys
associated as properties. This structure is designed to provide an
intuitive data model readily discoverable by a general astronomer.

Before deciding to use Data Central to host the Survey's data, the
SAMI Team worked on developing our own solution, samiDB
\citep{2015A&C....13...58K}. We developed this solution because, at
the time, there were no compelling options available to us for
organising and making public a data set such as ours. samiDB is
designed to require minimum setup and maintenance overhead while
providing a long-term stable format.  The solution also provides a
hierarchical organisation of the data, which has proved valuable as an
organisational model. The Team ultimately decided not to use samiDB to
present the data because Data Central offers ongoing support for the
data archive from the Australian Astronomical Observatory, and hence a
better chance that the Survey's data will remain generally and easily
available even after the Team has dissolved. However, the hierarchical
data model of samiDB has become a central part of the Data Central
design.

Further development of Data Central is planned. Most relevant to the
SAMI Galaxy Survey will be addition of all data products of the GAMA
Survey, enabling seamless querying of SAMI and GAMA as a
single data set. Also planned are more tools for interacting with the
data online. As this development progresses, the online user interface
is expected to continue to evolve, but the data of DR1 (and
their provenance), are stable and in their final form on the Data
Central service.

\section{Summary and Future}
\label{sec:future}

The SAMI Galaxy Survey is collecting optical integral-field
spectroscopy for $\sim$3,600 nearby galaxies to characterise the
spatially-resolved variation in galaxy properties as a function of
mass and environment. The Survey data are collected with the
Sydney/AAO Multi-object Integral-field Spectrograph (SAMI) instrument
on the Anglo-Australian Telescope. Survey targets are selected in two
distinct samples: a field sample drawn from the GAMA Survey fields,
and a cluster sample drawn from eight massive clusters.

With this paper, we release spectral cubes for 772 galaxies from the
GAMA sample of the Survey, one-fifth of the ultimate product. We also
release Value-Added products for the same galaxies, including maps of
emission-line fits, star-formation rate, and dust extinction. These
data are well suited to studies of the emission-line physics of
galaxies over a range of masses and rates of star formation. The
spectral cubes enable a multitude of science in other areas.

The next public data release of the SAMI Galaxy Survey is planned for
mid 2018, and will include further data and value-added products.

\hfill
\break

%\acknowledgments
\section*{Acknowledgements}
The SAMI Galaxy Survey is based on observations made at the
Anglo-Australian Telescope. The Sydney/AAO Multi-object Integral-field
spectrograph (SAMI) was developed jointly by the University of Sydney
and the Australian Astronomical Observatory. The SAMI input catalogue
is based on data taken from the Sloan Digital Sky Survey, the GAMA
Survey and the VST ATLAS Survey. The SAMI Galaxy Survey is funded by
the Australian Research Council Centre of Excellence for All-sky
Astrophysics (CAASTRO), through project number CE110001020, and other
participating institutions. The SAMI Galaxy Survey website is
\url{http://sami-survey.org/}.

JTA acknowledges the award of a SIEF John Stocker Fellowship.
MSO acknowledges the funding support from the Australian Research Council through a Future Fellowship (FT140100255).
BG is the recipient of an Australian Research Council Future Fellowship (FT140101202).
NS acknowledges support of a University of Sydney Postdoctoral Research Fellowship.
SB acknowledges the funding support from the Australian Research Council through a Future Fellowship (FT140101166).
JvdS is funded under Bland-Hawthorn's ARC Laureate Fellowship (FL140100278).
SMC acknowledges the support of an Australian Research Council Future Fellowship (FT100100457).
Support for AMM is provided by NASA through Hubble Fellowship grant \#HST-HF2-51377 awarded by the Space Telescope Science Institute, which is operated by the Association of Universities for Research in Astronomy, Inc., for NASA, under contract NAS5-26555.
CF gratefully acknowledges funding provided by the Australian Research Council's Discovery Projects (grants DP150104329 and DP170100603).
BC is the recipient of an Australian Research Council Future
Fellowship (FT120100660).

\section*{Contributions}
AWG and SMC oversaw DR1 and edited the paper.  SMC is the Survey's
Principal Investigator.  JBH and SMC wrote the introduction. JB
oversaw the target selection, and wrote those parts of the paper. NS
wrote sections on the changes to the data reduction, and oversaw the
data reduction with JTA and RS. ITH oversaw the emission line fits and
produced Figures~\ref{fig:vap-overview} and \ref{fig:multicom}. AMM
ran quality control on the emission line fits, produced the
higher-order value-added data products, and coordinated ingestion of
these to the database. BG helped coordinate preparation of value-added
products for release.  MJD and LC oversaw the formatting and
preparation of all data for inclusion in the online database. JvdS
prepared the survey overview diagram,
Figure~\ref{fig:poster-plot}. ADT and SMC measured the accuracy of the
WCS information and wrote the corresponding
Section~\ref{sec:wcs-accuracy}. RMM provided heliocentric velocity
corrections. FDE and JTA created Figures~\ref{fig:fwhm-change} and
\ref{fig:gal-mag} and contributed to the data reduction software and
to the assessment of the data quality, Section~\ref{sec:quality}. AWG,
EM, LH, SO, MV, KS, and AMH built the online database serving the
data.
Remaining authors contributed to overall Team operations including
target catalogue and observing preparation, instrument maintenance,
observing at the telescope, writing data reduction and analysis
software, managing various pieces of team infrastructure such as the
website and data storage systems, and innumerable other tasks critical
to the preparation and presentation of a large data set such as this DR1.

\bibliographystyle{andyapj}

\bibliography{bibliographies}

\appendix

\section{Aliasing  caused by differential atmospheric refraction correction and limited resolution and sampling}
\label{sec:subseeing}

The effects of differential atmospheric refraction can combine with
limited spatial resolution and incomplete sampling to introduce
aliasing into the spectra on scales comparable to the PSF. This
aliasing is not unique to IFS, though the generally poorer sampling in
both resolution and completeness tend to exacerbate the effect. We
will use the much simpler case of a long-slit spectrograph to explain
the effect.

To understand the impact of aliasing on spectral data in the presence
of differential atmospheric refraction, we consider a simple long-slit
image\footnote{For our purposes, a long-slit image is an image of a
  set of simultaneously observed spectra with spatial coordinate along
  the slit (chosen to be oriented along the paralactic angle for our
  examples) on the vertical axis and wavelength coordinates along the
  horizontal axis.} of a white continuum source (i.e.\ one with a flat
spectral-energy distribution in wavelength space). The slit has been
aligned with the parallactic angle so that atmospheric refraction acts
along the length of the slit. For illustrative purposes, we'll
consider the fairly extreme example of an object observed at a zenith
distance of 60 degrees. Throughout this section, we assume the seeing
is Gaussian, with one arcsecond FWHM.

\begin{figure*}
  \centering  
  \includegraphics[width=0.9\linewidth]{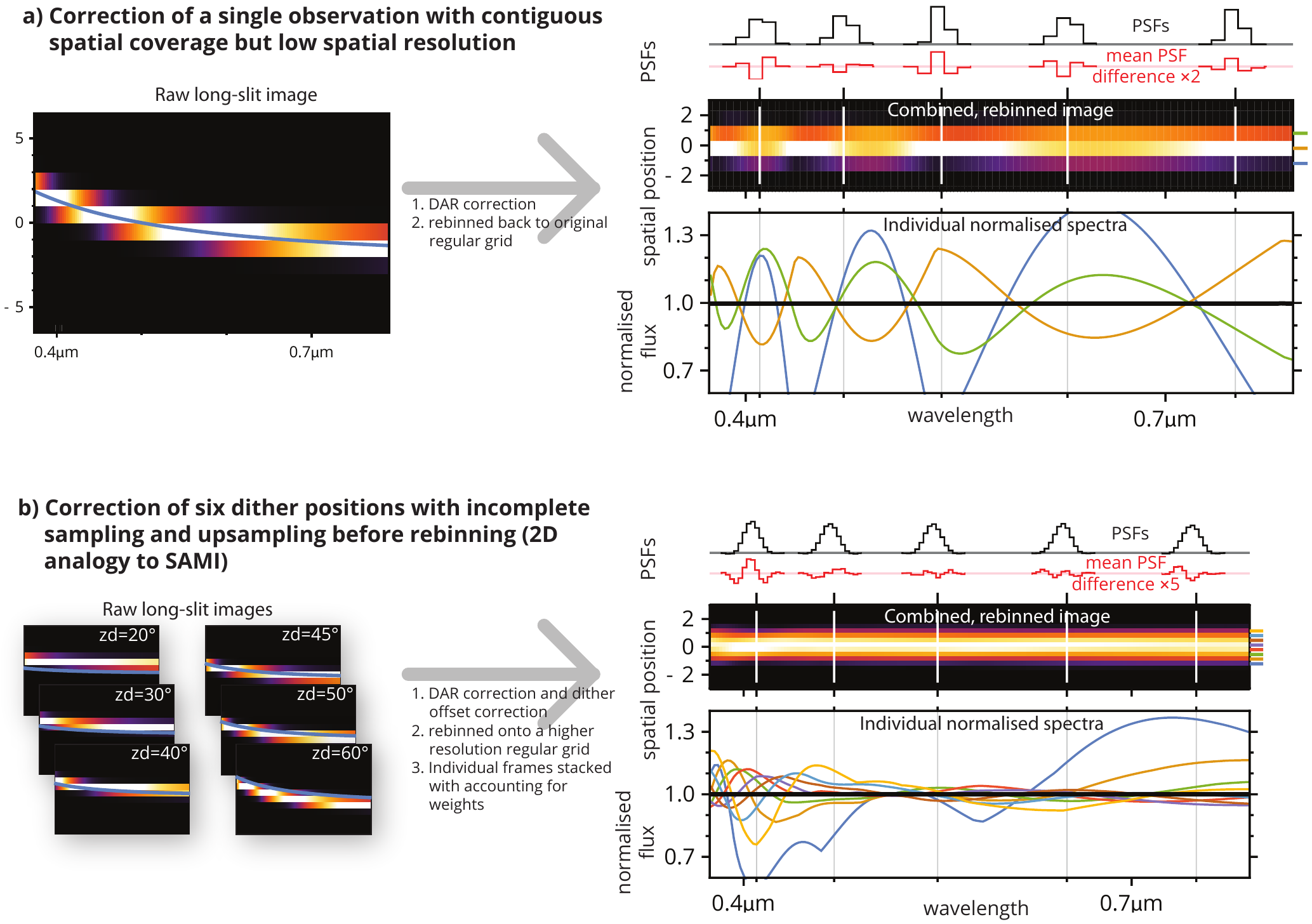}
  
  \caption[]{%
    \label{fig:dar-cartoon}%
    Long slit observations showing the effects of DAR and results of
    correcting for it in (a) a simple, single long-slit observation at
    high airmass, and (b) the combination of several long-slit
    observations at different air masses and dither positions---a
    close 2D analogy to our 3D data. The example shown in (a) is
    observed at a zenith distance of 60~degrees. The pixels are
    1~arcsec, and the underlying Gaussian PSF has a FWHM of
    1~arcsec. In (b), input frames are taken at six air-masses ranging
    from zenith distance (zd) of 20 to 60~degrees.  The underlying PSF
    is the same, and the spatial pixels are also 1~arcsecond, but the
    spatial sampling is incomplete. In the rebinning to correct for
    DAR, the data are up-sampled to 0.33~arcsec pixels before
    combining. \par
    \textbf{For each panel:} The left-hand side shows the raw
    long-slit image with a line showing the DAR at the centre of the
    slit over-plotted. The right-hand side shows the reconstructed
    long-slit image after DAR correction, including any
    rebinning. Vertical white lines mark the location of the spatial
    PSF shown above the image, with the difference from the mean PSF
    shown in red (scaled up to show detail). The plot below shows
    individual spectra from the image. The spatial location of each
    spectrum along the slit is shown by the corresponding coloured
    tick on the right of the image. These spectra have been normalised
    to highlight the relative differences in the spectra, which are
    entirely the result of the aliasing.  \par
    \emph{NOTE: Some PDF renders will attempt to smooth the pixels
      shown in this figure; we recommend using Acrobat Reader to see
      the actual, pixelated images as we intend.}%
  }
\end{figure*}

Consider a long-slit image of this object with a spatial scale of
one arcsecond per pixel. This image is shown (before correction for
DAR) on the left of Figure~\ref{fig:dar-cartoon}a. Note that the PSF, even
before correction, varies considerably along the wavelength axis due
to the poor spatial sampling of the data. A correction for DAR is
applied by shifting the pixels by the amount of the refraction along
the spatial direction and rebinning to the original regular
grid. After correction, the image of the object no longer shows a
position shift with wavelength (shown on the right in
Figure~\ref{fig:dar-cartoon}a). However, aliasing of the rebinning and
sampling are readily visible, causing the individual spectra at each spatial location (shown below the
image) to vary within the PSF, and the PSF (shown above the image) to
vary with wavelength.

Now, let us extend our example to be a close, 2D analogy to our own 3D
spectral cubes. This extended example is shown graphically in
Figure~\ref{fig:dar-cartoon}b. First, we observe the source at several
dither positions and air-masses. Second, we introduced gaps in the
spatial coverage that are smaller than and within the 1-arcsec pixels
(and therefore not readily apparent in the individual frames on the
left). The dithering ensures information falling in the gaps in one
frame will be picked up in another frame. It also tends to smooth out
the aliasing because individual dithers will each have a slightly
different aliasing PSF, which will be averaged out in the combination.
Finally, to bring our long-slit example closer to the actual process
used in SAMI, we add another complication: up-sampling. SAMI fibres
are 1.6~arcsec, but we sample the multiple observations onto a
0.5-arcsecond output grid.  Note that, in combining these six
individual frames, it is also necessary to track the weights of the
individual output pixels, which account for the gaps in the input
data. This extended example has all the same characteristics and
similar sampling dimensions of our actual SAMI data, except that we
are working with only one spatial dimension instead of two.

Reviewing the resulting combined, DAR-corrected long-slit image shows
that, despite its seeming smoothness, the PSF exhibits subtle but
important variations with wavelength and spatial position.  This
long-slit image is shown on the right of Figure
\ref{fig:dar-cartoon}b. The image is fairly smooth because up-sampling
and several dither positions and airmasses have averaged out some of
the aliasing. Yet the subtle differences in the PSF at different
wavelengths are still present. These differences are much more
apparent in the plot of individual spectra, where the spectrum at each
spatial position has been normalised to highlight the relative
differences. The spatial location of each of these spectra is shown by
the corresponding coloured tick on the right edge of the
image. Spectra further from the centre of the PSF (and with lower
total flux) tend to have larger relative deviations from the actual
spectral shape (this trend matches our analysis of observations of
individual stars with SAMI).

Pixelated (discretely sampled) data observed with DAR present show
effects of aliasing. These effects are exacerbated by poor spatial
resolution and incomplete sampling. Combining observations with many
dithers and different airmasses helps to average the aliasing
out. Up-sampling combined with sub-pixel dithering of the observations
can also reduce the severity of the aliasing. Aliasing is not
typically seen in long-slit data because the PSF is typically well
sampled. However, the tension in IFS between spatial sampling and
sensitivity, and the incomplete sampling present in many designs has
led to noticeable aliasing in IFS data. Although we have only
demonstrated the effect in 2D, long-slit data, DAR is only a 2D
effect, so our treatment of aliasing readily extends to 3D IFS data.

The general impact of aliasing is that the PSF varies both with
spatial and spectral position within either (2D) long-slit images or
(3D) spectral cubes. This effect is subtle, and in many cases can be
safely ignored without affecting results. There are, however, two
important exceptions. The first exception is cases where the PSF must
be known to very high accuracy. The second is when comparing data that
are widely separated in wavelength, for example emission-line ratios
or spatially resolved colours. Any analysis that averages over scales
larger than the PSF will not be affected by aliasing, such as measures
of radial gradients in galaxies and analysis that requires spatial
binning to bring out faint signals.

\end{document}